\documentclass{article}[12pt]
\usepackage{graphicx}
\usepackage{authblk}
\usepackage{amsmath}
\usepackage{amsfonts}
\usepackage{natbib}

\begin{document}
\newcommand{\beq}{\begin{equation}}
\newcommand{\eeq}{\end{equation}}
\newcommand{\pt}{\partial}

\title{Potential energy of atmospheric water vapor \\ and the air motions \\ induced by water vapor condensation\\ on different spatial scales}
\author[1]{A.~M.~Makarieva}
\author[1]{V.~G.~Gorshkov}
\affil[1]{Theoretical Physics Division, Petersburg Nuclear Physics Institute, Gatchina, St. Petersburg, Russia, elba@peterlink.ru}
\maketitle

\begin{abstract}
Basic physical principles are considered that are
responsible for the origin of dynamic air flow upon condensation of water vapor,
the partial pressure of which represents a store of potential energy in the atmosphere of Earth.
Quantitative characteristics of such flow are presented for several spatial scales.
It is shown that maximum condensation-induced velocities reach 160~m~s$^{-1}$ and are realized in compact
circulation patterns like tornadoes.
\end{abstract}

\large

\section{Introduction}

Atmospheric air on Earth conforms to the ideal gas law with a high accuracy. The main physical property of the ideal gas is that
its equation of state does not depend on molar masses of the gas mixture constituents.
At fixed temperature, a given value of pressure  can be obtained for a mixture with arbitrary molar masses
by setting the value of molar density of the gas.
The second important property of atmospheric air is the presence of a constituent that undergoes condensation
under terrestrial temperatures and pressures -- water
vapor.  Molar density of the moist air mixture is equal to the sum of molar densities
of the dry air and water vapor.

In a motionless atmosphere (with the impact of the greenhouse substances neglected) the air temperature would be the same
at all heights. Vertical distributions of all gases would follow the hydrostatic Boltzmann's distribution according
to their molar masses. The exponential scale height of water vapor with molar mass 18~g~mol$^{-1}$ would be about 13~km,
while the scale height of the major air constituents, nitrogen and oxygen, with similar molar masses $\sim 30$~g~mol$^{-1}$
would have scale heights of approximately 8~km. In this case water vapor would be saturated at the surface only
(due to the contact with the liquid hydrosphere) and had an undersaturated concentration elsewhere. There would be no evaporation
or condensation in such an atmosphere.

However, irrespective of the presence or absence of the greenhouse substances, such a hypothetical distribution of vertically isothermal moist air
appears to be unstable. Any fluctuation leading to an upward displacement of an air volume results in adiabatic cooling
of the rising air. Air temperature drops such that the equilibrium water vapor concentration dictated by Boltzmann's distribution
becomes oversaturated at all heights where the air ascends. This causes the water vapor to condense. Its concentration decreases
down to the saturated concentration. Condensation diminishes the total air pressure and disturbs Boltzmann's distribution of moist air.
The vertical gradient of air pressure becomes greater than the weight of a unit air volume.
There appears an upward-directed force acting on a unit air volume.
Static equilibrium of moist air
in the gravitational field is no longer possible. There appears a rising flow of air masses induced by condensation.
The process of condensation is sustained by continuous evaporation of water vapor from the hydrosphere.
The upward-directed force that acts on moist air causing it to rise adiabatically was termed the evaporative-condensational force
\citep{gm06,mg07,mg09a}.

Most part of the condensed water vapor leaves the atmosphere via precipitation. A minor part is maintained in the atmosphere
by the rising air flow. This imposes a drag force on the rising air flow and leads to a reduction of the vertical velocity.
However, unlike the flow of water vapor which condenses as it rises, the flow of the dry air components which conserved their mass,
cannot be unidimensional (vertical). There inevitably appear horizontal legs in the condensation-induced air circulation.
Condensation of water vapor in the ascending air produces both vertical
and horizontal pressure gradients. Therefore, the presence of water vapor in the atmosphere contacting
with a liquid hydrosphere leads to the formation
of three-dimensional circulation patterns.

In the following sections we derive relations between the vertical and horizontal components of air velocities and
condensation-induced pressure gradients. We further use the obtained results to describe large-scale circulation
with approximately constant velocities when the pressure gradient force and the turbulent friction force coincide.
We also apply these relations to describe hurricanes and tornadoes, where the pressure gradient forces appear to
significantly exceed the turbulent friction forces. So far the condensation-induced air motions have not received
a consideration in meteorology. Several
relevant observations regarding the conventional approaches are made in the footnotes to the main text.

\section{Continuity equation for moist air}

The equations of state for moist air as a whole, as well as for its components -- dry air
and water vapor, include one and the same universal molar gas constant $R$ and do not
depend on molar masses and mass densities of the components\footnote{Note that this fundamental universality of
ideal gas is masked in the meteorological literature by the common usage of mass densities $\rho = MN$,
$\rho_v = M_vN$ and $\rho_d = M_dN$ along with mass gass constants $R_{air} \equiv R/M$,  $R_v \equiv R/M_v$
and $R_d \equiv R/M_d$. Usually  $R_{air} \approx R_d$ is denoted as $R$, while the universal molar gas constant
is practically never used \citep[e.g.,][]{glick}.}:
\beq
\label{ig}
p = NRT, \,\,\, p_v = N_vRT,\,\,\, p_d = N_dRT,
\eeq
where $p$, $N$, $p_v$, $N_v$, $p_d$, $N_d$ are the pressure and molar density of moist air as a whole,
water vapor and dry air, respectively.

In a circulating atmosphere where no condensation takes place, the ratios of molar densities of all components
at all heights affected by the are equal to their mean atmospheric values. The reason is that
the diffusional velocities that would restore Boltzmann's distributions depending on molar densities and molar
masses of the components, are small compared to the dynamic velocities of the air flow. According to
observations, mixing ratios of the non-condensable air constituents and the molar mass of dry air are
the same at all heights in the troposphere.

In the absence of condensation the ratio $\gamma = N_v/N$ should not change at any changes of pressure
and temperature. Consequently, the process of condensation should be reflected in the changes of ratio $\gamma$.
The molar rate of condensation per unit volume (mole~m$^{-3}$~s$^{-1}$) that is
caused by the decrease of air temperature with height $z$ due to the adiabatic ascent of moist air with
vertical velocity $w$, is equal to
\beq
\label{cond}
wN\frac{\pt \gamma}{\pt z} = w \left(\frac{\pt N_v}{\pt z} - \frac{N_v}{N}\frac{\pt N}{\pt z}\right),
\eeq
\beq
\label{gam}
\gamma \equiv \frac{N_v}{N} = \frac{p_v}{p},\,\,\, \frac{d\gamma}{\gamma} = \frac{dN_v}{N_v} - \frac{dN}{N} = \frac{dp_v}{p_v} - \frac{dp}{p}.
\eeq

The meaning of Eq.~(\ref{cond}) is physically transparent. Condensation rate is determined by the change of
water vapor concentration minus the change of total air concentration that is not related to condensation.
We emphasize that the second term in brackets in Eq.~(\ref{cond})
comprises the relative change of molar density $N$ of moist air as a whole rather than molar density $N_d$ of its dry component.
This reflects the fact that restoration of equilibrium pressure distribution upon condensation affects the
air mixture as a whole, including the remaining water vapor.

Since the condensation rate (\ref{cond}) is a function of molar (not mass) densities, it is
convenient and physically transparent to write the continuity equation for ideal gas in terms of molar densities as well.
In Cartesian coordinates (assuming that there is no dependence of the flow on $y$) the continuity equation takes the form, see
Eq.~(\ref{cond}):
\beq
\label{cont}
\frac{\pt Nu}{\pt x} + \frac{\pt Nw}{\pt z} = w \left(\frac{\pt N_v}{\pt z} - \frac{N_v}{N}\frac{\pt N}{\pt z}\right).
\eeq
Taking into account that $N = N_d + N_v$ and
\beq
\label{contd}
\frac{\pt N_du}{\pt x} + \frac{\pt N_dw}{\pt z} = 0,
\eeq
we have from Eq.~(\ref{cont}):
\beq
\label{contv}
\frac{\pt N_vu}{\pt x} + \frac{\pt N_vw}{\pt z}  = w \left(\frac{\pt N_v}{\pt z} - \frac{N_v}{N}\frac{\pt N}{\pt z}\right).
\eeq
Expanding the derivatives in Eq.~(\ref{contv}) and multiplying both parts of the equation by $N/N_v$ we obtain
\beq
\label{prom}
u\frac{N}{N_v} \frac{\pt N_v}{\pt x} + w\frac{\pt N}{\pt z} + N\left(\frac{\pt u}{\pt x} + \frac{\pt w}{\pt z}\right) = 0.
\eeq
Now expanding the derivatives in the left hand part of Eq.~(\ref{cont}) and using Eq.~(\ref{prom}) we obtain
\beq
\label{prom2}
u \left(\frac{\pt N}{\pt x} - \frac{N}{N_v}\frac{\pt N_v}{\pt x}\right) = w \left(\frac{\pt N_v}{\pt z} - \frac{N_v}{N}\frac{\pt N}{\pt z}\right).
\eeq
Using definition (\ref{gam}) we can re-write the continuity equation (\ref{prom2}) as
\beq
\label{contg}
u\frac{\pt \ln\gamma}{\pt x} = - w\frac{\pt \gamma}{\pt z}.
\eeq
Now, turning from molar densities $N$ and $N_v$ to pressures $p$ and $p_v$ (\ref{ig}) we can see that
temperature dependencies cancel from Eq.~(\ref{contg}) and the latter becomes\footnote{Equation~(\ref{cont}) is equivalent
to the sum of Eqs.~(\ref{contd}) and (\ref{contv}). If there is no condensation, the right-hand part of
Eq.~(\ref{cont}) is zero. Equations~(\ref{cont}), (\ref{contd}) and (\ref{contv}) can then be re-written
in terms of mass densities $\rho = N/M$, $\rho_d = N_d/M_d$ and $\rho_v = N/M_v$. Here $M_v$ and $M_d$
are constant ($M_d$ is constant in agreement with observations because the molar ratios of the
dry air constituents do not change with height). But the molar mass of moist air $M = \gamma M_v + (1 - \gamma) M_d=
M_d (1 - 0.38\gamma)$ depends on $\gamma$ and, consequently, on $z$. Therefore, if the condensation rate
in the right-hand part of Eq.~(\ref{cont}) is not zero, it will change substantially upon transition
from molar to mass densities. Specifically, if condensation rate is written in form of the right hand part of Eq.~(\ref{cond})
with $N$ and $N_v$ changed to $\rho$ and $\rho_v$, respectively, this will lead to the appearance of an incorrect
multiplier $M_v/M$ in the right hand part of Eq.~(\ref{prom2}). This would contradict the physical meaning
of the ideal gas equations of state (\ref{ig}).}
\beq
\label{grp0}
u\left(\frac{\pt p}{\pt x} - \frac{1}{\gamma}\frac{\pt p_v}{\pt x}\right) = wp\gamma \left(\frac{1}{p_v}\frac{\pt p_v}{\pt z} -
\frac{1}{p} \frac{\pt p}{\pt z} \right).
\eeq
According to Clausius-Clapeyron equation, saturated pressure $p_v$ of water vapor depends on temperature only. When the
considered area is horizontally isothermal (temperature $T$ does not depend on $x$) we have $\pt p_v/\pt x = 0$.
This condition presumes the existence of an inflow of water vapor caused by local evaporation from the hydrosphere.
This inflow partly compensates condensation that occurs in the atmospheric column\footnote{It is
assumed in the conventional meteorology that the cause of atmospheric circulation is the
fact that the surface is not horizontally isothermal due to external differential heating. The condensation-induced
circulation, in contrast, arises on a horizontally isothermal surface (although this is not an indispensable condition).
Horizontal temperature inhomogeneities that can be observed after the circulation has established
are the consequences of the horizontal inhomogeneity of the process of water vapor condensation.}.
In this case Eq.~(\ref{grp0}) takes the form
\beq
\label{grp}
-\frac{\pt p}{\pt x} = \frac{w}{u}\frac{\Delta p}{h_\gamma},\,\,\,\Delta p \equiv p\gamma = p_v,\,\,\,h^{-1}_\gamma \equiv h^{-1}_v - h^{-1},
\eeq
\beq
\label{h}
h^{-1}_v \equiv -\frac{1}{p_v}\frac{\pt p_v}{\pt z},\,\,\,h^{-1} \equiv -\frac{1}{p}\frac{\pt p}{\pt z},\,\,\,\frac{\pt p_v}{\pt x}=0.
\eeq
All magnitudes entering (\ref{grp}) and (\ref{h}) depend on $x$ and $z$. Height $h_\gamma$ has the meaning
of characteristic height where all water vapor condenses. Heights $h_v$ and $h$ are the scale heights
of water vapor and moist air, respectively.

In a large-scale stationary circulation, where constant friction forces compensate the equally constant pressure gradient forces
\citep{mg09a},
horizontal velocity $u$ does not change with $x$. Velocity $w$ should be understood as the vertical velocity averaged over height $h_\gamma$.
The flux of air enters the circulation area horizontally via a vertical cross-section of area $Dh_\gamma$ and leaves the circulation area
vertically across a horizontal cross-section of area $DL$. Here $L$ is the horizontal dimension (length) of the circulation area counted
along the $x$-axis, $D$ is the circulation width counted along the $y$-axis perpendicular to the horizontal air flow. Taking
into account that the number of air mols $n_{in}$ that enter the circulation area differ from the number $n_{out}$ of air mols
leaving the circulation area by a relatively small number of mols of condensed water vapor, $n_{in} - n_{out} \sim \gamma n_{in} \ll
n_{in}$, to the accuracy of $\gamma \ll 1$ we can write (see also footnote 6 below):
\beq
\label{geo}
Dh_\gamma u = DLw,\,\,\,\frac{w}{u} = \frac{h_\gamma}{L}.
\eeq
Putting (\ref{geo}) into (\ref{grp}) we obtain
\beq
\label{grgeo}
-\frac{\pt p}{\pt x} = \frac{\Delta p}{L},\,\,\, \Delta p \equiv p\gamma = p_v.
\eeq
\noindent
Thus, total horizontal pressure drop is equal to partial pressure of water vapor.

To arrive to Eq.~(\ref{grgeo}) three equations have been used: Eq.~(\ref{contv}) for $N_v$,
equation $\pt p_v/\pt x = 0$, see (\ref{h}), that reflects that the circulation area is horizontally isothermal and that water vapor is saturated,
and Eq.~(\ref{contd}) for molar density $N_d$. The latter equation does not include the condensation rate term (\ref{cond}), which
is present in (\ref{cont}) and (\ref{contv}). Equation~(\ref{geo}) namely arises from Eq.~(\ref{contd}) for $N_d \gg N_v$.

It follows from Eq.~(\ref{grgeo}) that $p_v = \Delta p$ represents a store of potential energy,
which is converted to the kinetic energy of moving air masses as the water vapor condenses\footnote{
In the absence of friction at constant pressure gradient the horizontal velocity grows with distance $x$
as prescribed by Bernoulli's equation: $\rho u^2/2 = \Delta p = \gamma p$. Due to $\gamma \ll 1$ we observe
that $\Delta p$ is much smaller than $p$. On the other hand, we have $\Delta p/p = \gamma = \Delta \rho/\rho$.
Mass density $\rho$ changes little over the distance where air pressure changes by $\Delta p$: $\Delta \rho = \gamma \rho \ll \rho$.
From this in the meteorological literature it is concluded that changes in density $\rho$ due
to condensation can be neglected \citep[e.g.,][]{sa08}. . The resulting physical inconsistency is masked by the fact that changes of $\rho$
and the formation of $\Delta p$ are conventionally ascribed to independent physical causes. In particular, it is
common in circulation models to take pressure profiles from observations. The relative smallness of horizontal
pressure drop $\Delta p$ is never discussed.

But if we put $\Delta \rho/\rho = \Delta p/p$ equal to zero, no velocity can form and the circulation cannot exist.
This statement is general and does not depend on why $\Delta p$ and $\Delta \rho = \rho_0 -\rho$ actually form. Mathematically,
the error can be spotted as follows. From the equation of state (\ref{ig}) we have $p = C\rho$, where $C \equiv RT/M = {\rm const.}$
for the considered horizontally isothermal surface. From Bernoulli's equation we then have $u^2 = 2\Delta p/\rho =
2C\Delta \rho/\rho = 2C (\Delta \rho/\rho_0) (1+\Delta \rho/\rho_0 + ...)$.
As one can see,
discarding $\Delta \rho$ compared to $\rho$ does indeed correspond to discarding the term of a higher order of smallness.
But with respect to the pressure gradient, the main effect is proportional $\Delta \rho$, which
is the term of the first order of smallness.
If the smallness is set to zero, the effect disappears.}.
By integrating (\ref{grgeo}) we obtain the following expression for potential energy $P(x) = p(x)$
that is defined to the accuracy of a constant term:
\beq
\label{P}
P(x) = p(x) = p(0) + \frac{\Delta p}{L}x.
\eeq
In the incoming air flow at $x = L$ water vapor is present everywhere in that part of the atmospheric column
that is affected by the circulation. At $x = 0$ this water vapor
has been completely used up and is partially replaced by locally evaporated water vapor.

\begin{figure}
\begin{center}
\includegraphics[width=0.9\textwidth]{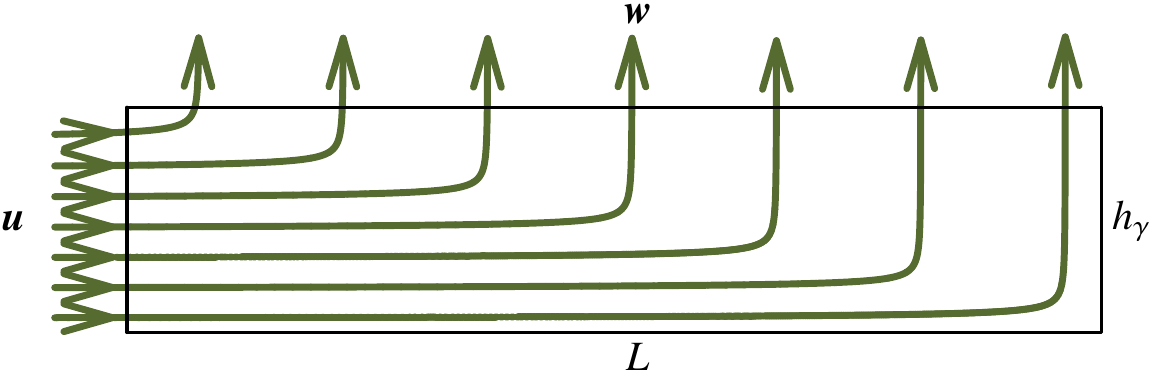}
\end{center}
\caption{Air streamlines for a large-scale circulation of length $L$ and height $h_\gamma$. The continuity equation (\ref{geo})
is $uh_\gamma = wL$.}
\end{figure}

Strictly speaking, Eqs.~(\ref{geo}) and (\ref{grgeo}) are valid for a bunch of streamlines that enter
the circulation area in the horizontal direction and leave it in the vertical direction. The number
of horizontal streamlines present in the atmospheric column at a given $x$ decreases as one travels
inside the area at the expense of those streamlines that have left the area in the vertical direction.
Velocitites $u$ and $w$ (\ref{geo}) and pressure gradient (\ref{grgeo}) are the same for each streamline;
the mean horizontal velocity in the column linearly decreases, while the height of ascent linearly grows, Fig.~1.
All the above forms of the continuity equation, Eqs.~(\ref{grp})-(\ref{grgeo}), describe the kinematics
of the air flow at given values of $u$ and $w$. These velocities should be determined from Euler's
equations with an account of friction\footnote{The major friction force per unit area of the Earth's surface, which
opposes the pressure gradient force that would otherwise accelerate the air, is the friction force
that can be called gravitational, as it is proportional to the weight of atmospheric column $\mu\rho g h =
\mu p_s$, $\mu = z_T/h \sim 10^{-5}$. Here $p_s$ is surface pressure, $z_T \sim 0.1$~m is the surface roughness
(it is proportional to the height of vegetation cover, oceanic waves etc.), $h \sim 10$~km is the scale height
of the atmosphere.
The same form of gravitational friction, $\mu \rho g h$, is due to frictional dissipation of
liquid drops precipitating or suspended in the atmosphere. In this case $\mu$ has the meaning of the relative volume occupied by the drops,
$\mu \sim \gamma (w/w_b) \sim 10^{-5}$, where $w_b$ is the mean downward velocity of the drops, $w/w_b \sim 10^{-3}$.
The gravitational friction force does not depend on velocities $u$ or $w$
and can be represented as $\mu\rho g h = \rho u_{g*}^2$, where $\mu = u_{g*}^2/gh$ is Froude's number,
$u_{g*}$ has the meaning of rotation velocity for the turbulent eddies that bud from the main air flow
due to gravitational friction. The gravitational friction force $\rho u_{g*}^2$ exceeds by 30 times
the force of aerodynamic friction $c_D\rho u^2 = \rho u_*^2$ ($u_{g*}^2 \sim 30 u_{*}^2$)
that is usually taken into account in the Navier-Stokes equations and the formulation of Reynolds stress \citep{mg09a}.}.
Thus, condensation occurs as the moist air
ascends in the vertical direction and is maintained due to evaporation from the horizontal Earth's surface.
The associated condensational-evaporative force induces makes the moist air masses circulate along the streamlines
that include both vertical and horizontal regions.

\section{Condensation in the adiabatically ascending air}
According to Clausius-Clapeyron equation, we have
\beq
\label{CC}
\begin{split}
\frac{dp_v}{p_v} = \xi \frac{dT}{T},\,\,\, p_v(T) = p_{v0}\exp (\xi_0 - \xi),\\
-\frac{1}{p_v}\frac{\pt p_v}{\pt z} = \xi \frac{\Gamma}{T} \equiv h_v^{-1},\,\,\,\Gamma \equiv -\frac{\pt T}{\pt z},\,\,\,\xi \equiv \frac{L_v}{RT},
\,\,\,\xi_0 \equiv \frac{L_v}{RT_0}.
\end{split}
\eeq
\noindent
where $L_v = 45$~kJ~mol$^{-1}$ is the molar heat of vaporization (latent heat).

Moist air obeys the hydrostatic equilibrium distribution\footnote{Equation~(\ref{hd}) should be more appropriately referred to as
the equation of {\it aerodynamic equilibrium in the gravitational field of Earth}. Indeed, this equilibrium does not correspond to a static Boltzmann's distribution of gases
with different molar masses. It arises in the result of the air ascent which occurs with a sufficiently high velocity $w$, which
is the same for all gases (including the remaining water vapor) despite their different molar masses. In the system of rest of the ascending
air, Eq.~(\ref{hd}) corresponds to a hydrostatic distribution of air with a constant molar mass $M$ and departs
from the latter only insignificantly due to the fact that $\gamma \ll 1$ decreases with height.
Vertical turbulent flux of water vapor associated with evaporation from the surface leads to the fact that
the vertical velocity of water vapor $w_v$ in the ascending air flow is always larger than the vertical velocity of air $w$,
$w_v = w + \Delta w_v$ \citep{mg07}. At $\gamma \ll 1$ this does not change the flux of moist air as a whole, $Nw = N_d w + N_vw_v = Nw(1+\gamma\Delta w_v/w)
\approx Nw$. The ratio between vertical velocities of water vapor and air as a whole reaches its maximum value $\Delta w_v/w \approx
1$ in a large-scale circulation where $w$ is small. In compact intense circulation where condensation rate greatly
exceeds the local rate of evaporation, we have $\Delta w_v \ll w$ and $w_v \approx w$.
}
\beq
\label{hd}
-\frac{1}{p}\frac{\pt p}{\pt z} = h^{-1} = \frac{Mg}{RT},\,\,\,M = M_d (1 - 0.38 \gamma).
\eeq
Functions $\Gamma(z)$, $T(z)$ and $\gamma(z)$ can be found
from the first law of thermodynamics under condition that the ascent is adiabatic,
using the definition of $\gamma$ (\ref{gam}) and the Clausius-Clapeyron equation (\ref{CC}):
\beq
\label{1LT}
c_p\frac{\pt T}{\pt z} - \frac{1}{N}\frac{\pt p}{\pt z} + L_v \frac{\pt \gamma}{\pt z} = 0;\,\,
-\frac{1}{\gamma}\frac{\pt \gamma}{\pt z} = - \frac{1}{p_v}\frac{\pt p_v}{\pt z} + \frac{1}{p}\frac{\pt p}{\pt z} = h_v^{-1} - h^{-1}
\equiv \frac{1}{h_\gamma}.
\eeq
The first equation in (\ref{1LT}) is the first law of thermodynamics for an adiabatic process, with the third term describing
condensation, see (\ref{cond}); $c_p = c_v + R = (7/2) R$ is the molar
heat capacity at constant pressure.

System of equations (\ref{CC})-(\ref{1LT}) can be re-written in a closed form:
\beq
\label{cls}
-\frac{\pt T}{\pt z} \equiv \Gamma(z)=\Gamma_d \frac{1+\gamma \xi}{1+\mu\gamma\xi^2}(1-0.38\gamma),
\eeq
\beq
\label{cls2}
-\frac{1}{\gamma}\frac{\pt \gamma}{\pt z} = h_{\gamma}^{-1} = \xi \frac{\Gamma(z)}{T} - \frac{M_d(1-0.38\gamma)g}{RT}.
\eeq
$$\Gamma_d \equiv \mu\frac{T}{h_d} = 9.8\,{\rm K\,km}^{-1},\,
h_d \equiv \frac{RT}{M_dg},\, \mu \equiv \frac{R}{c_p}.$$
\noindent
Here $\Gamma_d$ is the value of $\Gamma(z)$ at $\gamma =0$, i.e. it is the adiabatic lapse rate of air temperature for dry
air\footnote{In the meteorological literature one can find statements \citep[e.g.,][p.~S12436]{poe09} that the release of
latent heat $L_v$ upon condensation warms the air, as well as that latent heat is a major source of energy for some
types of circulation \citep[e.g., see discussion and references in][]{mgln10}.
In reality, according to the Clausius-Clapeyron equation, condensation occurs when there is
an external cause of cooling (e.g., adiabatic ascent). Condensation cannot warm the air to a temperature higher
than the air had prior to condensation. During condensaton, the water vapor concentration decreases, which,
according to the Clausius-Clapeyron law, corresponds to a drop (not a rise) of air temperature.}.
Eq.~(\ref{cls}) at $\gamma \ll 1$ coincides
with the well-known expression for moist adiabatic lapse rate of air temperature \citep{glick}\footnote{
We emphasize that $\gamma$ in Eqs.~(\ref{cls}), (\ref{cls2}) is defined as in Eq.~(\ref{gam}). At $p_d \to 0$
we have $\gamma \to 1$ and $\pt \gamma/\pt z \to 0$. In the meteorological literature formulae~(\ref{cls}) and (\ref{1LT})
are written for $\gamma_d \equiv p_v/p_d$ instead of for $\gamma \equiv p_v/p$ (\ref{gam}). At $\gamma \ll 1$
we have $\gamma \approx \gamma_d$, but at $p_d \to 0$ we have $\gamma_d \to \infty$, which leads to a physically
meaningless expression $\pt \gamma_d/\pt z \to \infty$ if one replaces
$\gamma$ by $\gamma_d$ in the continuity equation (\ref{contg}) and Eqs.~(\ref{cls2}) and (\ref{cond}).}.
Solutions of Eqs.~(\ref{cls}), (\ref{cls2}) are shown in Fig.~2.

\begin{figure}
\begin{center}
\includegraphics[width=0.9\textwidth]{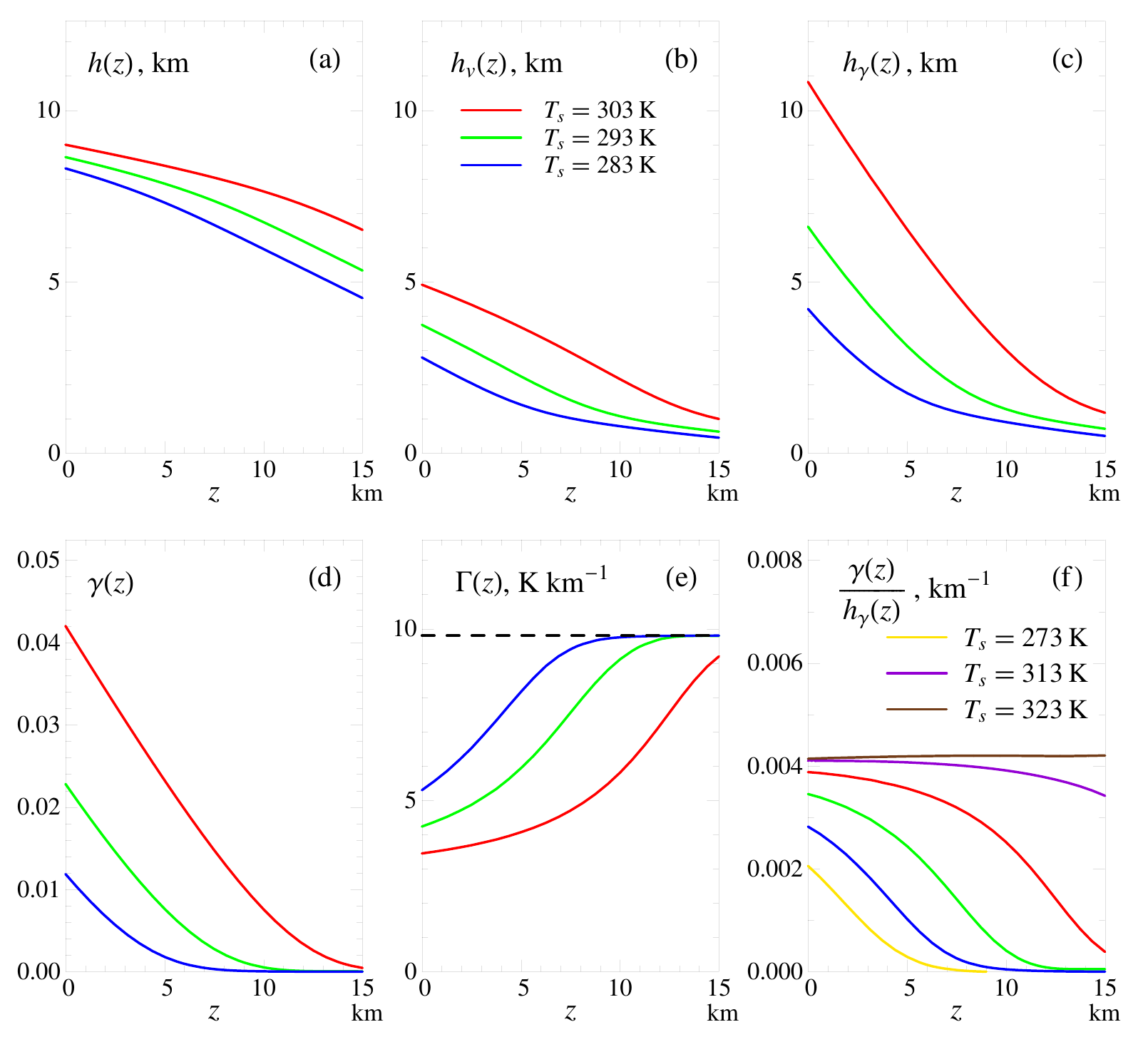}
\end{center}
\caption{Dependence on height $z$ of the major distribution functions of moist air
for several different values of surface temperature $T_s$ satisfying the system of equations (\ref{cls})-(\ref{cls2}).\newline
(a): $\displaystyle h(z)^{-1} = -\frac{1}{p}\frac{\pt p}{\pt z}$ is the scale height ~(\ref{hd}) of the moist air pressure;\newline
(b): $\displaystyle h_v(z)^{-1} = -\frac{1}{p_v}\frac{\pt p_v}{\pt z}$ is the scale height~(\ref{CC}) of the partial pressure
of saturated water vapor;\newline
(c): $\displaystyle h_\gamma \equiv -\frac{1}{\gamma}\frac{\pt \gamma}{\pt z}$ is the scale height of water vapor condensation (\ref{cls2});
$p$ and $p_v$ are pressures of moist air and saturated water vapor, respectively.\newline
(d): $\gamma(z) \equiv p_v(z)/p(z)$ is the mixing ratio of saturated water vapor;\newline
(e): $\Gamma(z) = -\pt T/\pt z$ is the moist adiabatic lapse rate of air temperature, the dashed line
indicates the dry adiabatic lapse rate $\Gamma_d = 9.8$~K~km$^{-1}$;\newline
(f): $\gamma(z)/h_\gamma(z) = -\pt\gamma/\pt z$ is the relative intensity of water vapor condensation.}
\end{figure}

Due to the large value of the ratio $L_v/R \equiv T_v \approx 5300$~K, the dimensionless ratio $\xi \equiv T_v/T$
is always much larger than unity. Therefore, at $\gamma \to 1$ (this would happen if
at constant temperature the dry component were largely removed from the atmosphere
and the water vapor partial pressure became the dominant contributor to total air pressure) the adiabatic lapse rate
(\ref{cls}) ceases to depend on $z$ and tends to $\Gamma \to (\Gamma_d)(0.62/\mu\xi) = 1.2$~K~km$^{-1}$. Scale height $h_v$ (\ref{h})
of saturated water vapor tends to the hydrostatic equilibrium value $h_v = T/(\Gamma \xi) \to h_d/0.62 = h_{v\,stat} = RT/(M_vg) = 13.5$~km. The
right-hand part of the last equality in Eq.~(\ref{1LT}) turns to zero; $\gamma$ ceases to depend on $z$.
The third term containing $\pt \gamma/\pt z$ in the equation of the second law of thermodynamics (\ref{1LT}) vanishes.
Condensation, evaporation and the condensation-induced air circulation all stop.

The transition to hydrostatic equilibrium during adiabatic ascent of moist air at $\gamma \to 1$ implies
that the weight of saturated water vapor column of unit area at any height becomes equal to saturated vapor
pressure at this height: the evaporative-condensational force disappears. Consequently, the adiabatic
ascent of moist air can no longer be maintained by this force and cannot arise spontaneously. In this
case, in contrast to the case of $\gamma \ll 1$, the stable stationary state of the atmosphere is not
the state of self-sustained air circulation, but the the state of hydrostatic equilibrium of motionless air (water vapor)
with constant air temperature at all heights.

The condensation-induced
circulation of the modern atmosphere occurs at a global mean surface temperature that is optimal for life. This temperature
fixes the saturated concentration of water vapor in accordance with the Clausius-Clapeyron equation (\ref{CC}).
Therefore, the condensation-induced circulation only becomes possible due to the high concentration of atmospheric nitrogen,
which can be called an air "ballast". Indeed, unlike CO$_2$, nitrogen is not a greenhouse gas; unlike oxygen, gaseous nitrogen
is chemically inert and cannot lead to excessive oxidation
or fires in the biosphere. The major role of atmospheric nitrogen consists in making the value of $\gamma$ small at a fixed temperature.
The condensation-induced circulation makes the biotic pump of atmospheric moisture \citep{mg07} and a hydrological cycle
on land possible, thus ensuring that land is habitable for life. This suggests that
the existing concentration of atmospheric nitrogen should have been formed by the time while life started colonizing the land
and the continental forest cover developed.

Note that if $\gamma \to 1$ due to increasing temperature (and not due to the removal of the dry air component),
then the exponential growth of $p_v = \gamma p$ in accordance with the Clausius-Clapeyron equation~(\ref{CC})
makes the evaporative-condensational force grow exponentially as well, such that the condensation-induced
atmospheric circulation is ensured at any small share of the dry component in the total air pressure, see (\ref{h}), (\ref{grgeo})
and Fig.~2f.

For a given value of $\gamma(0) \equiv \gamma_s \ll 1$ at the surface, $\gamma(z)$ declines rapidly with increasing $z$,
Fig.~2d. Accordingly, the adiabatic temperature lapse rate $\Gamma(z)$ (\ref{cls}) rises from its minimal value at the surface
approaching the dry adiabatic lapse rate $\Gamma_d$ at large heights, Fig.~2e. Any circulation pattern includes areas where
the air masses rise (in the region where evaporation and condensation are more intense) and areas where
the air masses descend (in the region where evaporation and condensation are less intense). In the region of descent
the air masses warm adiabatically while descending, so condensation cannot occur. Consequently, neglecting horizontal mixing, the
adiabatic temperature lapse rate in the region of descent should be equal to the dry adiabatic lapse rate.
Air temperature $T$ and air pressure $p$ related to temperature by the equation of state (\ref{ig}) decrease with height
more slowly in the region of ascent than they do in the region of descent. Pressure difference $\Delta p$ between the
regions of ascent and descent has a negative value at the surface, then decreases by absolute magnitude with growing
height, approaches zero and changes its sign at a certain height $z_c \sim h_\gamma$, where horizontal
velocity $u$ changes its direction \citep{mgsnl10}.

The equation of state (\ref{ig}) can be written in the following form
\beq
\label{ig2}
p = \rho g h,\,\,\,\rho = NM, \,\,\, h = RT/(Mg).
\eeq
This relationship for $p$ does not in reality depend on either $M$ or $g$, which actually cancel
in~(\ref{ig2}). But written in this form, presure has a simple physical meaning of potential energy
in the gravity field, where $h$ represents the height of atmospheric column, $\rho g$ represents
weight of a unit air volume in the air column, while $p$ represents the total weight of the
air column of a unit area. Relationshp~(\ref{ig2}) holds for any height $z$, with $\rho$ and $T$ depending on $z$.
At the Earth's surface $z = 0$; surface values of all variables are denoted by low index $s$. The difference
in surface weights between two hypothetical static air columns -- in one of which saturated water vapor has condensed
at all heights, while in the other no condensation took place -- is equal to the difference of the right-hand parts of
the first equality in (\ref{ig2}) written for the two columns:
\beq
\label{diff}
p_{vs}gh_{vs} - p_{vs}gh_{s} = p_{vs}g (h_{vs} - h_s) < 0.
\eeq
\noindent
Here surface pressure $p_{vs}$ of water vapor is set to be saturated, so its value is determined
by the surface temperature and is the same in the two columns considered. It follows from~(\ref{diff})
that the first column, where condensation took place, cannot remain in static equilibrium, as surface
air pressure appears to be larger than the column weight. The surplus of pressure at the surface
causes air to rise and all water vapor to ultimately condense as the air rises and cools. This illustrates
the physical role of water vapor contained in the air column as a store of potential energy
available for atmospheric circulation.

The global mean store of potential energy per unit atmospheric water vapor mass is of the order
of $\overline{p_{vs}/\rho_{vs}} \approx R\overline{T_s}/M_v \sim 1.3 \times 10^5$~J~(kg~H$_2$O)$^{-1}$
at the global mean surface temperature $\overline{T_s} \approx 288$~K. The global mean precipitation rate is
$\Pi \sim 10^3$~kg~H$_2$O~m$^{-2}$~year$^{-1}$ \citep{lv}. Thus, the global rate of potential energy release associated
with water vapor condensation is of the order of $\Pi R\overline{T_s}/M_v \sim 4$~W~m$^{-2}$. This potential energy flux is sufficient
to drive the general atmospheric circulation of Earth. The power of the latter has been estimated
at around $\sim 1\%$ of the global power of $2.4 \times 10^2$~W~m$^{-2}$ of the absorbed solar radiation \citep{lor67}.

\section{Large-scale and compact circulations}

In a stationary circulation condensation of water vapor in the region of adiabatic ascent of air masses
must be compensated by an inflow of water vapor from the hydrosphere to the atmosphere via evaporation.
This follows from the general continuity equation written in the form of (\ref{prom2})
or (\ref{contg}). If $\gamma$ does not change in the horizontal direction, the change of $\gamma$
in the vertical direction that is due to condensation, also turns to zero.

\begin{figure}
\begin{center}
\includegraphics[width=0.8\textwidth]{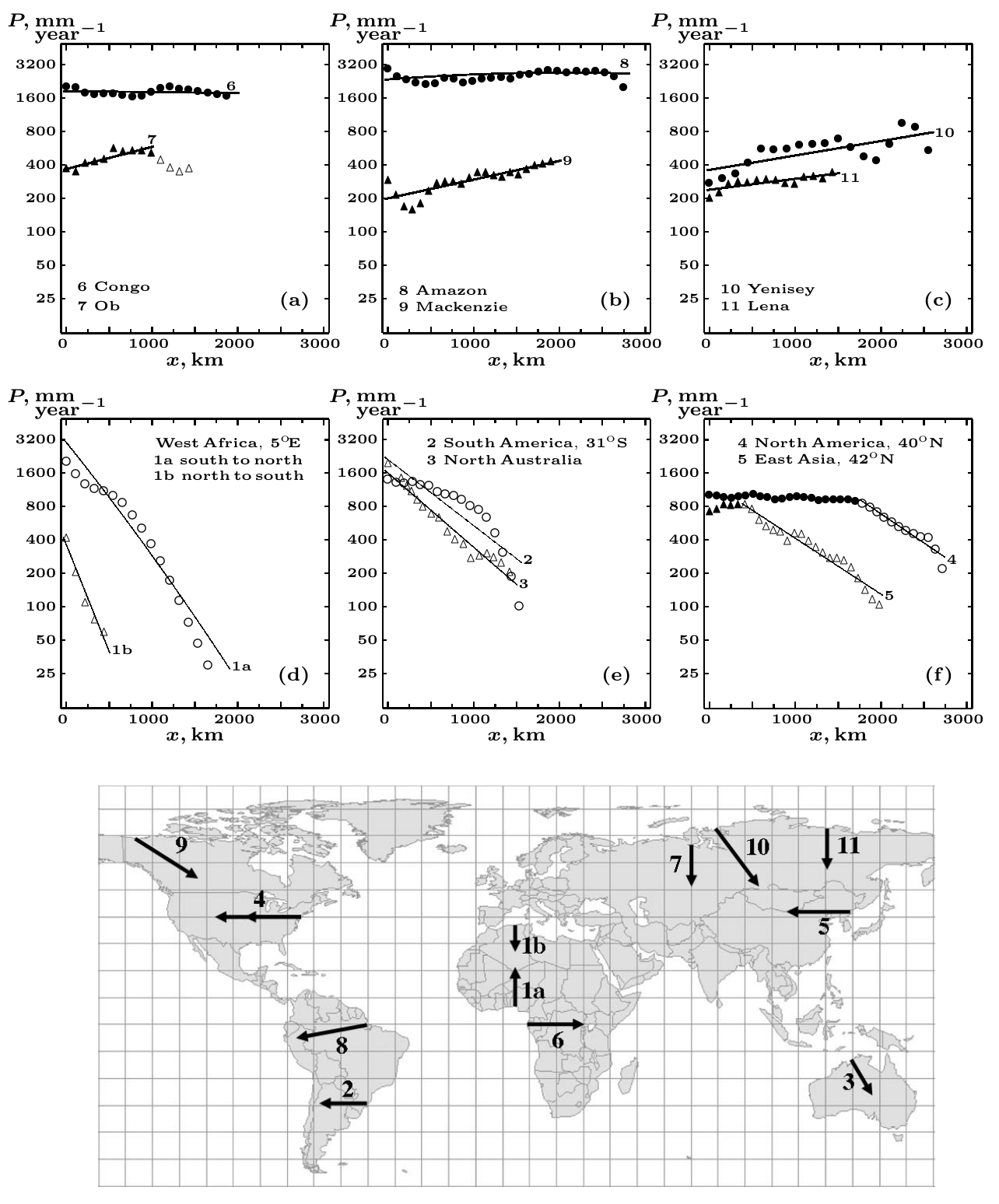}
\end{center}
\vspace{-0.5cm}
\caption{Biotic pump of atmospheric moisture in river basins covered by natural forests
(black symbols) as compared to disturbed and unforested regions (open symbols). Shown is the
mean annual precipitation on land associated with large-scale regional atmospheric circulation
patterns as dependent on distance from the ocean. Arrows on the map indicate which regions were considered.\newline
(a), (b), (c): data for the world's largest river basin covered by natural forests.
In the basins of world's largest rivers  -- Amazon and Congo -- precipitation is twice higher than over
the ocean and does not decrease with growing distance from the ocean.
Precipitation in the basins of the northern rivers that flow from the south
grow with distance from the ocean proportionally to the increase in solar radiation \citep{mgl09}.
In the Ob basin precipitation declines in the deforested region near the place where Ob is joined
by Irtysh.
\newline
(d), (e), (f): precipitation over non-forested areas decline exponentially with distance
from the ocean. In panel (f) temperate forests of the North America maintain nearly constant precipitation
until the deforested region is reached (line 4).}
\end{figure}

In a large-scale circulation where pressure gradient forces are compensated by friction forces such that
horizontal velocity $u$ does not change, the maintenance of a constant value of water vapor partial pressure
in the horizontally moving air masses, $\pt p_v/\pt x = 0$, is achieved via continuous local evaporation.
Stipulation $\pt p_v/\pt x = 0$ describes circulation that is isothermal in the horizontal direction.
Evaporation flux should be of the same order of magnitude as the condensation flux. The latter exceeds
the evaporation flux by no more than approximately twofold -- at the expense of horizontal import
of water vapor evaporated within the region of descent where condensation does not take place. The relative
value of evaporation dictates the direction of air circulation: the air flow in the lower atmosphere is directed
towards the region of higher evaporation, which entails more intense condensation.

Namely this pattern determines how the biotic pump of atmospheric moisture functions. Evaporation from the forest canopy of
natural forests can exceed evaporation from the open oceanic surface by over twofold. This leads to the appearance
of atmospheric flow bringing moisture from the ocean to land to compensate for the continental moisture loss due to
gravitational runoff
\citep{gm06,mg06,mg07,mgl09,mg09a}, Fig.~3.
Intense condensation over forest canopy creates a region of low pressure of continental scale
that sustains the ocean-to-land air flow.
The regular and persistent atmospheric moisture delivery
of the biotic pump becomes possible due to the fact that natural forests control the processes of evaporation
(via transpiration and intercept) and condensation (via production and emission of biogenic condensation nuclei).
This regular moisture transport does not allow spatial and temporal fluctuations of condensation to develop
thus preventing the formation of hurricanes and tornadoes over the natural forest canopy spread
over millions of square kilometers. All these biotic pump features disappear in deforested regions,
that, via the transitional monsoon-type regime of intermittent drafts and floods, ultimately turn
to deserts that are deprived of precipitation altogether, Fig.~3.

In a compact circulation, friction forces are negligibly small compared to pressure gradient forces. This makes it
possible for the air masses to accelerate and achieve catactrophic velocities observed within hurricanes and tornadoes.
In this case condensation rate can exceed the local evaporation rate as supported by solar radiation
by several orders of magnitude. The circulation pattern can remain stationary if only, after the local store
of water vapor is depleted, the pressure field moves along to the adjacent area where water vapor is still abundant.
Having depleted the water vapor in the new area, the pressure field moves to the next area and consumes water vapor there,
and so on \citep{vg90,vg95,vg00}. Thus, in this case the import of water vapor to the
circulation area occurs dynamically via movement of the pressure field, while the consumption of water vapor occurs thermodynamically
in the course of adiabatic ascent of air masses.

Hurricane energetics is determined by conservation of the sum of kinetic energy of the radial, tangential and vertical air movement
and potential energy of condensing water vapor on each streamline, as described by the Bernoulli integral. (While Euler's equations are non-linear,
Bernoulli's integrals that represent the sum of kinetic and potential energy, are linear. Therefore, any bunch of streamlines
(sum of particular streamlines) is also a Bernoulli integral.) As friction is negligible, angular momentum is conserved along each streamline.
The apparent non-conservation of angular momentum within the hurricane is caused by a superposition of different streamlines that
correspond to different boundary conditions (see Section~8).

Re-distribution of energy between the radial, tangential and vertical velocity components of the air flow results in the appearance
of an eye of hurricane or tornado near the condensation center. Due to angular momentum conservation, kinetic energy of the tangential flow grows
more rapidly towards the center than does kinetic energy corresponding to the radial and vertical velocity. Therefore, all energy is ultimately converted
to the energy of the tangential air flow, while the radial and vertical flow components cease to exist (their energy becomes zero).
As far as condensation rate
and the associated potential energy are proportional to vertical velocity, condensational potential energy turns to zero as well.
All the energy store of the Bernoulli integral is now concentrated in the kinetic energy of the tangential air flow. The rotating
air flow cannot move further towards the center because of the zero radial velocity, so a spot of calm weather is formed in the center.
In the absence of considerable friction, the kinetic energy of the tangential air flow can persist for a long time
endangering any object that can become a source of frictional dissipation. We summarize that the properties of hurricanes
and tornadoes follow from (1) the properties of the potential energy associated with condensation, (2) conservation of angular momentum,
(3) conservation of air flow (continuity equation) with a high precision, and cannot be understood without taking the corresponding
physical processes into account.

\section{Condensational pressure gradient in a radially symmetrical circulation}

Let us now consider the structure of hurricanes and tornadoes in greater quantitative detail. Such compact circulations
have a conspicuous condensation center and can be considered in the cylindrical system. The continuity equation in the cylindrical coordinates
under the assumption of radial symmetry (no dependence on angle $\varphi$) have the following form, cf. (\ref{cont})-(\ref{contv}):
\beq
\label{contr}
\frac{1}{r}\frac{\pt Nur}{\pt r} + \frac{\pt Nw}{\pt z} = w \left(\frac{\pt N_v}{\pt z} - \frac{N_v}{N}\frac{\pt N}{\pt z}\right).
\eeq
\beq
\label{contdr}
\frac{1}{r}\frac{\pt N_dur}{\pt r} + \frac{\pt N_dw}{\pt z} = 0,
\eeq
\beq
\label{contvr}
\frac{1}{r}\frac{\pt N_vur}{\pt r} + \frac{\pt N_vw}{\pt z}  = w \left(\frac{\pt N_v}{\pt z} - \frac{N_v}{N}\frac{\pt N}{\pt z}\right).
\eeq
\noindent
Here $u$ and $w$ are the radial and vertical velocities of the air flow. Due to radial symmetry, tangential velocity $v$
does not enter (\ref{contr})-(\ref{contvr}). Expanding the derivatives in Eq.~(\ref{contvr}) and
multiplying both parts of the equation by $N/N_v$ we obtain
\beq
\label{promr}
u\frac{N}{N_v} \frac{\pt N_v}{\pt r} + w\frac{\pt N}{\pt z} + N\left(\frac{1}{r}\frac{\pt ur}{\pt r} + \frac{\pt w}{\pt z}\right) = 0.
\eeq
Now expanding the derivatives in the left hand part of Eq.~(\ref{contr}) and using Eq.~(\ref{promr}) we obtain
the following continuity equation where condensation is taken into account:
\beq
\label{prom2r}
u \left(\frac{\pt N}{\pt r} - \frac{N}{N_v}\frac{\pt N_v}{\pt r}\right) = w \left(\frac{\pt N_v}{\pt z} - \frac{N_v}{N}\frac{\pt N}{\pt z}\right).
\eeq
Using the equation of state (\ref{ig}) we finally have
\beq
\label{grp0r}
u\left(\frac{\pt p}{\pt r} - \frac{1}{\gamma}\frac{\pt p_v}{\pt r}\right) = wp\gamma \left(\frac{1}{p_v}\frac{\pt p_v}{\pt z} -
\frac{1}{p} \frac{\pt p}{\pt z} \right)
\eeq
or
\beq
\label{contgr}
u\frac{\pt \ln\gamma}{\pt r} = - w\frac{\pt \gamma}{\pt z}.
\eeq
When the condensation area is horizontally isothermal, we have $\pt p_v/\pt r = 0$, so Eq.~(\ref{grp0r}) takes the form
similar to (\ref{grp})-(\ref{h}):
\beq
\label{grpr}
-\frac{\pt p}{\pt r} = \frac{w}{u}\frac{\Delta p}{h_\gamma},\,\,\,\Delta p \equiv p\gamma = p_v,\,\,\,h^{-1}_\gamma \equiv h^{-1}_v - h^{-1},
\eeq
\beq
\label{hr}
h^{-1}_v \equiv -\frac{1}{p_v}\frac{\pt p_v}{\pt z},\,\,\,h^{-1} \equiv -\frac{1}{p}\frac{\pt p}{\pt z},
\,\,\,\frac{\pt p_v}{\pt r} = 0,
\eeq
\noindent
where scale height $h_v$ and $h$ are determined from the conditions of adiabatic ascent and hydrostatic equilibrium for moist air
including the remaining non-condensed water vapor, see Section~3. As in (\ref{grp})-(\ref{h}), $h_\gamma$ has the meaning
of the scale height where most part of water vapor condenses. All variables entering (\ref{grpr}) and (\ref{hr}) depend
on $r$ and $z$.

To find the dependence of pressure gradient (\ref{grpr}) on $r$ and $z$ let us use the continuity equation in the integral form.
Air flow converging towards the condensation center penetrates via vertical round wall of radius $r$, circumference $2\pi r$
and height $h_\gamma$ and leaves the condensation area via horizontal disk of area $\pi r^2$:
\beq
\label{conti}
2\pi r h_\gamma u N = 2\pi \int_0^rN_d wrdr,\,\,\,{\rm or}\,\,\,\frac{1}{r}\frac{\pt Nur}{\pt r} = N_dw.
\eeq
\noindent
Here we took into account that all water vapor that entered the circumfurence has condensed and it is only the dry air component
that leaves the condensation area. Recalling the smallness of $\gamma \ll 1$ and considering that the relative pressure
difference between the hurricane center and the outskirts does not exceed 10\%, one can neglect the corresponding
changes of $N \approx N_d$ in (\ref{conti}). Then we have:
\beq
\label{contiv}
\frac{w}{h_\gamma} = \frac{1}{r} \frac{\pt ur}{\pt r} = \frac{u}{r} + \frac{\pt u}{\pt r}.
\eeq
Putting (\ref{contiv}) into (\ref{grpr}) we obtain
\beq
\label{p}
-\frac{\pt p}{\pt r} = \frac{\Delta p}{ur} \frac{\pt ur}{\pt r} = \Delta p\frac{\pt \ln ur}{\pt r},
\,\,\,\Delta p \equiv p\gamma \equiv \frac{1}{2}\rho u_c^2,
\eeq
\noindent
where $p$ and $\gamma$ are the corresponding values at the surface or, more accurately, at the height of
boundary layer where relative humidity reaches unity and condensation commences; $u_c$ is the
condensational velocity, which is the velocity scale that determines the magnitude of kinetic energy to which potential
energy $\Delta p$ can be converted.

\section{Profiles of pressure and velocities in hurricanes and tornadoes}

Thus, potential energy $P(r)$ that is responsible for the formation of hurricanes and tornadoes,
depends on radial velocity $u(r)$ (the latter itself appears due to conversion of potential energy
$P(r)$ to kinetic energy) and is equal to
\beq
\label{Pr}
P(r) = p(r) = \Delta p \ln ur + p(r_p).
\eeq
The Bernoulli integral of Euler equation for a streamline has the form
\beq
\label{BI}
\Phi(r) \equiv \frac{1}{2}\rho_s(u^2 + v^2 + w^2) + \Delta p \ln ur = \Phi(r_p),
\eeq
\noindent
where $r_p$ is a fixed radius, $v$ is tangential velocity that is perpendicular to radius $r$.
As a boundary condition to fix $r_p$ it is natural to set $u(r_p) \sim 5$~m~s$^{-1}$
to be equal to the average wind velocity outside the area of the considered compact
circulation. Conservation of angular momentum constrains the dependence
of $v$ on $r$. Using the continuity equation (\ref{contiv}) that relates $w$ and $u$ to each other, we then have
\beq
\label{am}
vr = v_pr_p,\,\,\,v = v_p\frac{r_p}{r},\,\,\,\frac{w}{h_\gamma} = \frac{1}{r} \frac{\pt ur}{\pt r}.
\eeq

We now go over to the following dimensionless variables:
\beq
\label{var1}
x \equiv \frac{r}{r_p},\,\,\Delta p \equiv \frac{1}{2}\rho u_c^2,\,\,u_c \equiv \frac{2\Delta p}{\rho} = \gamma\frac{2p}{\rho},
\eeq
\beq
\label{var2}
u \to \frac{u}{u_c},\,\,v\to \frac{v}{u_c},\,\,w \to \frac{w}{u_c},
\,\,\frac{\pt p}{\pt r} \to \frac{1}{\Delta p}\frac{\pt p}{\pt x},\,\,p \to \frac{p}{\Delta p},
\eeq
\beq
\label{var3}
\frac{v_p}{u_c} = \frac{A_p}{A_c} \equiv a,\,\,\,v_p \equiv v(r_p),\,\,\,A_p \equiv v_pr_p,\,\,\,A_c \equiv u_cr_p.
\eeq
\noindent
Here $u_c$ is the condensational velocity scale (\ref{p}), $A_c$ is the condensational angular momentum,
$a$ represents the dimensionless angular momentum $A = A_p$
(per unit mass density $\rho$) that is conserved.

Leaving the pressure and velocity notations unchanged for the dimensionless variables
(which is equivalent to choosing the units $u_c$ and $\Delta p$ of velocity and pressure
measurements equal to unity)
we obtain for (\ref{BI}):
\beq
\label{BIa}
\Phi(x) \equiv u^2 + v^2 + w^2 + \ln ux = \Phi(1),
\eeq
\beq
\label{BIb}
v = \frac{a}{x},\,\,\,w = \beta\frac{1}{x}\frac{\pt ux}{\pt x},\,\,\,\beta \equiv \frac{h_\gamma}{r_p},
\,\,\,p(x) = \ln ux + p(1).
\eeq
Putting (\ref{BIb}) into (\ref{BIa}) we finally obtain the following equation on radial velocity $u$:
\beq
\label{u}
\Phi(x) \equiv u^2 + \frac{a^2 + \beta^2 u^2}{x^2} + 2\beta^2 \frac{u}{x}\frac{\pt u}{\pt x} + \beta^2\left(\frac{\pt u}{\pt x}\right)^2 + \ln ux = \Phi(1).
\eeq

Due to the Earth's rotation, in the inertial frame of reference the air volumes rotate with an angular velocity $\omega = \Omega \sin \vartheta$, where
$\vartheta$ is the latitude angle where circulation takes place, $\Omega = 2\pi/\tau$, $\tau = 24$~h. Far from the condensation area due
to friction effects the rotation is similar to solid body rotation. There is no radial convergence of the air flows towards the condensation center.
Angular momentum is proportional to $\omega r^2$ and declines proportionally to $r^2$ with decreasing $r$. Radial velocity and air convergence
towards the condensation center arise at $r = r_p$, where the pressure gradient forces associated with the on-going condensation significantly
exceed the friction forces. Starting from $r < r_p$ angular momentum in the inertial frame proportional to $\omega r_p^2 = v_p r_p = vr$,
$v_p = \omega r_p$, is conserved. Tangential velocity increases towards the center in accordance to (\ref{am}) as a consequence of the
radial symmetry of the condensation-induced
pressure gradient forces and the smallness of friction forces that can be put approximately equal to zero at $r < r_p$.

In the system of observations, due to friction between the air and the surface,
the air is motionless at $r > r_p$ and the angular momentum is equal to zero. At $r < r_p$ there appears
a non-zero radial velocity and the air masses start to converge towards the center.
At this moment the Coriolis force that is perpendicular to radial velocity vector and to the vector of
angular velocity is non-central; it starts to curl the converging air masses, which leads to the increment of tangential velocity
and angular momentum. This occurs until tangential velocity becomes considerably larger than radial velocity. Then the Coriolis force
becomes a central force, and the angular momentum is further conserved.

Tangential velocities in the inertial frame of reference $v$ ($v = v_pr_p/r = \omega r_p^2/r$) and in the system of
observations $v_{ob}$ are related as follows:
\beq
\label{vob}
v_{ob} = v - \omega r = \omega \left(\frac{r_p^2}{r} - r \right) = v_p \left(\frac{1}{x} - x\right) = v_p \frac{1-x^2}{x},\,\,\,
x \equiv \frac{r}{r_p}.
\eeq
\noindent
Equation~(\ref{vob}) describes the fact that in the system of observations from the value of tangential velocity in the inertial frame,
where angular momentum is conserved, one substracts tangential velocity equal to the velocity of a solid body rotation $\omega r$.
Angular momentum per unit mass density $\rho$ in the system of observations is proportional to $v_{ob}r$:
\beq
\label{A}
A_{ob} = v_{ob}r = \omega r_p^2\left(1-\frac{r^2}{r_p^2}\right) = A(1-x^2),
\eeq
\noindent
where $A = \omega r_p^2 = v_p r_p$ is the angular momentum that is conserved in the inertial system. Change of angular momentum
in the system of observations as the air moves radially towards the center is determined by Coriolis force:
\beq
\label{Cor}
\begin{split}
\frac{d\textit{\textbf{A}}_{ob}}{dt} =  -2 [\textit{\textbf{r}}[\textit{\textbf{\boldmath{$\omega$}V}}(\textit{\textbf{r}})]]=
-2\textit{\textbf{\boldmath{$\omega$}}}ur,\,\,\,\textit{\textbf{V}} = \textit{\textbf{u}}+\textit{\textbf{v}}+\textit{\textbf{r}},
\,\,\,(\textit{\textbf{V}}\textit{\textbf{r}}) = ur,\\
A_{ob}(r) =\int_{r_p}^r \frac{dA}{dt}dt = \int_{r_p}^r \frac{dA}{dt}\frac{dt}{dr}dr = -\int_{r_p}^r 2\omega ur\frac{1}{u}dr =\\
= \omega r_p^2\left(1-\frac{r^2}{r_p^2}\right) = A(1-x^2),\\
u = \frac{dr}{dt},\,\,\,A_{ob}(r_p) = 0,
\end{split}
\eeq
\noindent
which coincides with Eq.~(\ref{vob}). Therefore, at $x \ll 1$ ($r \ll r_p$) the angular momenta in the system of observations and
the inertial system are conserved and coincide. Due to this fact it is convenient to consider hurricanes and tornadoes
in the inertial system where the Coriolis force is absent. Thus, the conserved dimensionless angular
momentum $a$ is defined as follows, see~(\ref{var3}):
\beq
\label{vpa}
a \equiv \frac{r_p\omega }{u_c},\,\,\,a_{ob}=a(1-x^2)\approx a,\,\,\,x \ll 1.
\eeq
An important peculiarity of a compact circulation like hurricanes and tornadoes is the fact that in order
for the circulation to remain stationary (independent of time), it is necessary that
the entire circulation pattern moves as a whole with velocity $U$ along the Earth's surface.
Within each streamline the air becomes deprived of water vapor after it ascends in the condensation area.
When all the air in the condensation area ascends to height $h_\gamma$, the circulation pattern has to
move over a horizontal distance of the order of the circulation size $r_p$. Consequently,
the horizontal velocity of the system movement is $U \sim \overline{w}r_p/h_\gamma$, where $\overline{w}$
is the mean vertical velocity of the ascending air within the circulation pattern.

In hurricanes the vertical velocity is small, so $\beta \ll 1$, see~(\ref{BIb}), and can be put equal to zero in~(\ref{u}).
Then Eq.~(\ref{u}) assumes a simple form:
\beq
\label{us}
u^2 +\frac{a^2}{x^2}+ \ln\left(\frac{u}{u_1}x\right) = a^2 + u_1^2,\,\,\,p(x) = \ln\left(\frac{u}{u_1}x\right),\,\,\,u_1\equiv u(1).
\eeq
Vertical velocity $w$ is related to radial velocity $u$ by Eq.~(\ref{BIb}). Point of maximum $u = u_m$ that
is obtained from $\pt u/\pt x = 0$ corresponds to $x = x_m = \sqrt{2}a$. At smaller $x < x_m$ the value
of $u^2$ rapidly declines with decreasing $x$. Contribution of the logarithmic term $-\ln(u/u_1)$
vanishes as $u$ decreases down to $u_1 < u_m$. Further increment of $-\ln(u/u_1)$ at $u < u_1$
does not have a physical meaning. Indeed, since the water vapor is depleted in the end of the streamline at some small $x$,
the condensational potential turns to zero. We thus assume that $x = x_0$, where radial velocity decreases from its
maximum value to $u = u_0 = u_1$, is the point where the condensational pressure gradient ceases to act (see Section 9). At this
point the radial and vertical velocities as determined by condensation within the considered area are effectively equal to zero.
All condensational energy is concentrated in the kinetic energy that corresponds to the tangential velocity
of the air flow.

\section{Rotation of the eye}
According to Eq.~(\ref{us}), the value of $x_0$ at which $u = u_0 = u_1$ is a root of the following equation, Fig.~4:
\beq
\label{x0}
-\ln x_0 = \frac{a^2}{x_0^2}(1-x_0^2) \approx \frac{a^2}{x_0^2},\,\,\,x_0^2 \ll 1;\,\,\,x_0 = x_0(a).
\eeq
\noindent
Point $x_0$ determines the radius of the hurricane eye for the case when the air in the eye were motionless
with a zero kinetic energy of rotation. However, in this case at $x = x_0$ the tangential velocity reaches its maximum
value $v_0 = a/x_0 = \omega_0x_0$. Therefore, the air in the eye cannot be motionless being in contact with the rotating windwall
at $x = x_0$.
In the absence of friction losses between the eye and the windwall, the eye should rotate as well with a constant angular
velocity $\omega_0$ determined by the tangential velocity of the windwall at $x = x_0$. Energy for the rotation of the eye
can be only borrowed from the kinetic energy of the rotating windwall, because at $x = x_0$ the radial and vertical velocities,
as well as the potential energy of condensation, are close to zero. The energy for eye rotation is therefore a certain share of
the full potential energy of water vapor condensation that was converted to the kinetic energy of the
windwall\footnote{The
pressure gradient force is perpendicular to isobars. Coriolis force is perpendicular to the velocity vector and is proportional
in magnitude to velocity. If the two forces act in the opposite directions and compensate each other, one can determine
the constant value of velocity corresponding to movement along the isobars. In the inertial frame of reference for the case of radially symmetrical closed isobars
this corresponds to the equality between the centrifugal force that depends on constant tangential velocity $v$ and the pressure gradient force:
$v^2/r + \pt p/\pt r = 0$ (the so-called cyclostrophic balance). If there is no friction, the angular momentum is conserved,
and the rotational movement with zero radial velocity and constant tangential velocity can continue infinitely,
similarly to satellite rotation on a terrestrial orbit.
Friction decreases tangential velocity and, hence, the centrifugal force; this disturbes the balance of forces and causes
some radial movement towards the center of the isobars. In the existing theories of hurricane formation the pressure gradient
force is approximated from the condition of approximate cyclostrophic balance using the dependence of tangential velocity on radius and the
fact that the observed radial and vertical velocities are small compared to the former. Friction is considered as the major
cause of the convergence of air masses towards the hurricane center \citep[e.g.,][]{sm08}. The cause-and-effect link
between the radial convergence and the existence of the radial pressure gradient is neglected. However,
nature works the other way round: it is namely the radial convergence of moist air with a non-zero radial velocity $u$ and the
physically inseparable ascent of air masses
with vertical velocity $w$, see (\ref{contiv}), that determine the pressure fall from the outer environment towards the hurricane center, see~(\ref{p}).
If there is no radial convergence and the radial velocity is zero, then there is no ascent of moist air, no condensation of water vapor,
no condensation-induced pressure gradient, no close isobars and no approximate cyclostrophic balance --  in short, no circulation and, consequently,
no friction. This is the major difference of the condensational potential from the velocity-independent gravitational potential that keeps
the rotating satellite on its orbit.
The condensational potential energy as the air converges towards the center is mainly converted to the kinetic energy of
rotation determined by tangential velocity $v$. This results in
the development of huge velocities namely due to the absence of friction. Friction can only lead to dissipation of this energy impeding the
air acceleration and preventing the development
of maximum possible velocities. }$^,$\footnote{The appearance of eye rotation
due to the contact of the eye with the rotating windwall represents a re-distribution of the ordered kinetic energy. It cannot
not be considered as turbulent friction losses (i.e., energy losses on the formation of the chaotic turbulent air flows).}.

\begin{figure}
\begin{center}
\includegraphics[width=0.9\textwidth]{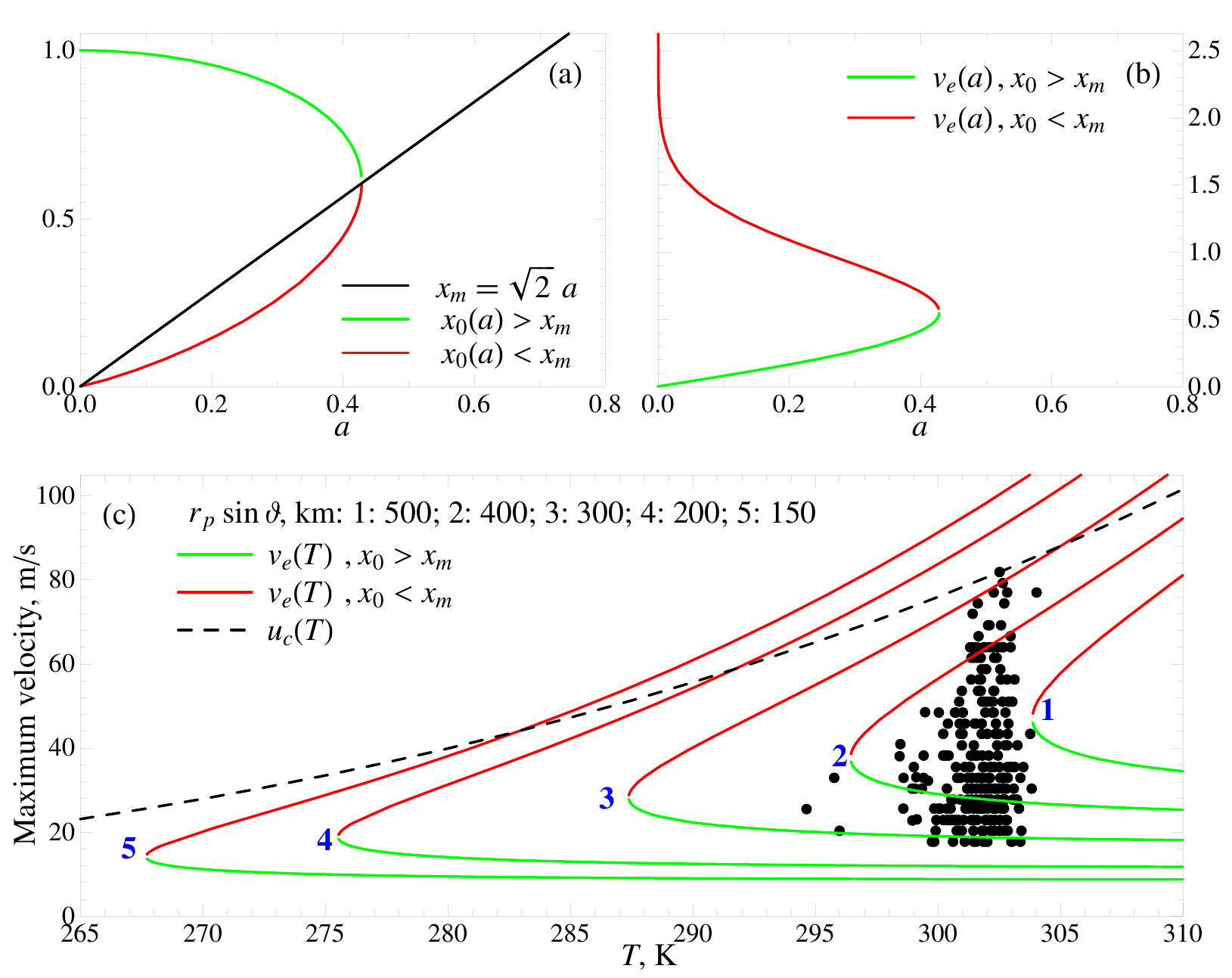}
\end{center}
\caption{The regime of hurricane existence corresponds to the two solution branches of equation~(\ref{x0}). The green curves
in all panels correspond to the branch where the eye radius $x_0(a)$ grows up to $x_0 = \sqrt{2}a = 0.6$ as $a$ increases
up to $a \approx 0.43$. The red curves in all panels correspond to the branch where the eye radius grows
from $x_0 = 0.6$ to $x_0 = 1$ with $a$ diminishing from $0.43$ to zero.\newline
\noindent
(a): Two branches of the solution of the approximate equation~(\ref{x0}).\newline
\noindent
(b): Maximum dimensionless tangential velocity $v_e(a) = a/x_e = a/(1.28 x_0)$ (\ref{xe}). The branch
with lower velocity corresponds to widening of the eye accompanied by the decrease of condensational energy.
\newline
\noindent
(c): Absolute values of maximum tangential velocities $v_e(a) = u_ca/x_e$ depending on temperature
for five different values of parameter $r_p\sin \vartheta$. For example, $r_p\sin \vartheta = 300$~km
corresponds to a streamline with the outer radius $r_p = 600$~km formed at latitude $\vartheta = 30^{\rm o}$.
The value of $u_c$ grows exponentially with increasing temperature as $u_c = [2\gamma(T)RT/M]^{1/2}$,
see (\ref{var1}), $\gamma(T) \equiv p_v(T)/p$, $p_v(T)$ conforms to the Clausius-Clapeyron equation
$dp_v/p_v = (L/RT)dT/T$, see (\ref{CC}), $\gamma = 0.042$ at $T = 303$~K.\newline
\noindent
Black dots represent the data from Fig.~3 of \citet{mi06} that
correspond to maximum wind velocities observed in 270 tropical cyclones.}
\end{figure}

Transfer of kinetic energy to the eye reduces the maximum tangential velocity of the hurricane that is observed within
the windwall from $v_0$ to $v_e = a/x_e$. The eye radius expands from $x_0$ to $x_e > x_0$. The air in the eye
rotates with a constant angular velocity $\omega_e = v_e/x_e < \omega_0$ and tangential velocity $v = \omega_e x$, $x < x_e$.
For the definitiveness sake, below we continue to call $x_0$ as the radius of the (motionless) eye while
referring to $x_e$ as to the radius of the windwall (of the rotating eye).
According to the energy conservation law, the kinetic energy of eye rotation should be equal to the kinetic energy of
rotation of the windwall segment $x_0 \le x \le x_e$ with tangential velocity $v = a/x$ which would
take place if the eye remained motionless, see Fig. 5a,b.
Neglecting the air density change in the eye as compared to the windwall (see footnote 5) and
cancelling
the common multiplier $2\pi$, this stipulation can be written in the following form, see Fig.~5a:
\beq
\label{seg}
\omega_e^2 \int_0^{x_e} x^2 x dx = a^2 \int_{x_0}^{x_e}\frac{xdx}{x^2},\,\,\,\omega_e^2 = \frac{a^2}{x_e^4}.
\eeq
\noindent
Performing the integration and cancelling the common multiplier $a^2$ we obtain
\beq
\label{xe}
\ln\frac{x_e}{x_0} = \frac{1}{4},\,\frac{x_e}{x_0} =  e^{0.25} = 1.28;\,x_e(a) = 1.28x_0(a),
\,v_e(a) = \frac{a}{1.28 x_0(a)}.
\eeq
Tangential velocity $v$ grows with decreasing $x$ at $x > x_e$ and declines with decreasing $x$ at $x < x_e$ as follows, see (\ref{vpa}):
 \beq
\label{v2}
v = \begin{cases} {\displaystyle \frac{a}{x}}, & x \ge x_e, \, {\displaystyle vx = a =\frac{\omega r_p}{u_c} = {\rm const.},} \, \omega = \Omega \sin\vartheta,\\
\,\\
{\displaystyle \frac{ax}{x_e^2}}, & x \le x_e, \, {\displaystyle vx = \frac{ax^2}{x_e^2} \ne {\rm const.}} \end{cases}
\eeq
The dependence of the maximum value of tangential velocity $v_e = v(x_e)$ on $a$ is shown in Fig.~4b.

Stationary existence of the rotating windwall and the rotating eye modifies pressure profile $p(x)$ at $x \le x_e$
in such a manner that the pressure gradient force and the centrifugal force within the eye coincide.
In order that an additonal pressure drop could form within the eye where no condensation takes place, some part of the air must
be exported away from the eye outside the hurricane area ($x > 1$) during the eye formation in such a manner that
the parts of the pressure profile inside and outside the rotating eye joined smoothly and featured no discontinuity at the windwall $x = x_e$.
This condition can be written as \citep{mg09b}
\beq
\label{sh1}
\frac{\pt p}{\pt x} = 2\frac{v^2}{x} = 2\omega_e^2x,\,\,\,p(x) = \omega_e^2x^2 + p(0),\,\,\,\omega_e = \frac{a}{x_e^2},\,\,\,x \le x_e;
\eeq
\beq
\label{sh2}
p(x) = \ln\left(\frac{u}{u_1}x\right),\,\,\,p(1) = 0,\,\,\,x \ge x_e.
\eeq
(Multiplier "2" in (\ref{sh1}) arises from the equation $\pt p/\pt r = v^2/r$ as one goes to the dimensionless
variables (\ref{var2}): dividing both parts of the equality by $\Delta p = \rho_s u_c^2/2$ and changing
to the dimensionless velocity $v/u_c$.) In Eq.~(\ref{sh2}) the condensational potential is set
equal to zero outside the condensation area $x \ge 1$ ($r \ge r_p$). Thus, $p(x)$ describes how air pressure declines from its normal
value it has outside the hurricane. Accordingly, $-p(0) = \delta p$ is equal to the total (dimensionless) pressure fall within
the hurricane including the pressure fall associated with the eye rotation.
In ordinary units, the pressure fall is equal to $\delta p \Delta p = \gamma p \delta p$.
Combining (\ref{sh1}) and (\ref{sh2})
we have, see Fig.~5c,d:
\beq
\label{p2}
p(x) = \begin{cases} {\displaystyle \ln\left(\frac{u}{u_1}x\right)}, & x \ge x_e \\ \,\\
{\displaystyle a^2\frac{(x^2-x_e^2)}{x_e^4}+\ln\left(\frac{u}{u_1}x\right)}, & x \le x_e \end{cases},
\eeq
\beq
\label{dp}
\delta p = \frac{a^2}{x_e^2}-\ln\left(\frac{u_e}{u_1}x_e\right).
\eeq
\noindent
The first and second terms in (\ref{dp}) represent the pressure fall within the eye and outside the eye, respectively.
At $a = 0.12$ ($r_p = 400$~km, $\vartheta =20^{\rm o}$, $u_c = 83$~m~s$^{-1}$ at $\gamma \approx 0.04$)
we have $x_0 \approx 0.12$, see~Fig.~4a; $x_e \approx 0.15$, see (\ref{xe});
$u_e/u_1 \sim 2$, see Fig.~5b. Thus, from Eq.~(\ref{dp}) we obtain
$\delta p \approx 3$, Fig.~5d. For the absolute pressure fall we have
$\delta p\Delta p/p \sim 3\gamma\sim 0.13$ at $\gamma \approx 0.04$, which means that the air pressure in the eye
center is at maximum 13\% lower than the air pressure outside the hurricane.

\begin{figure}
\begin{center}
\includegraphics[width=0.8\textwidth]{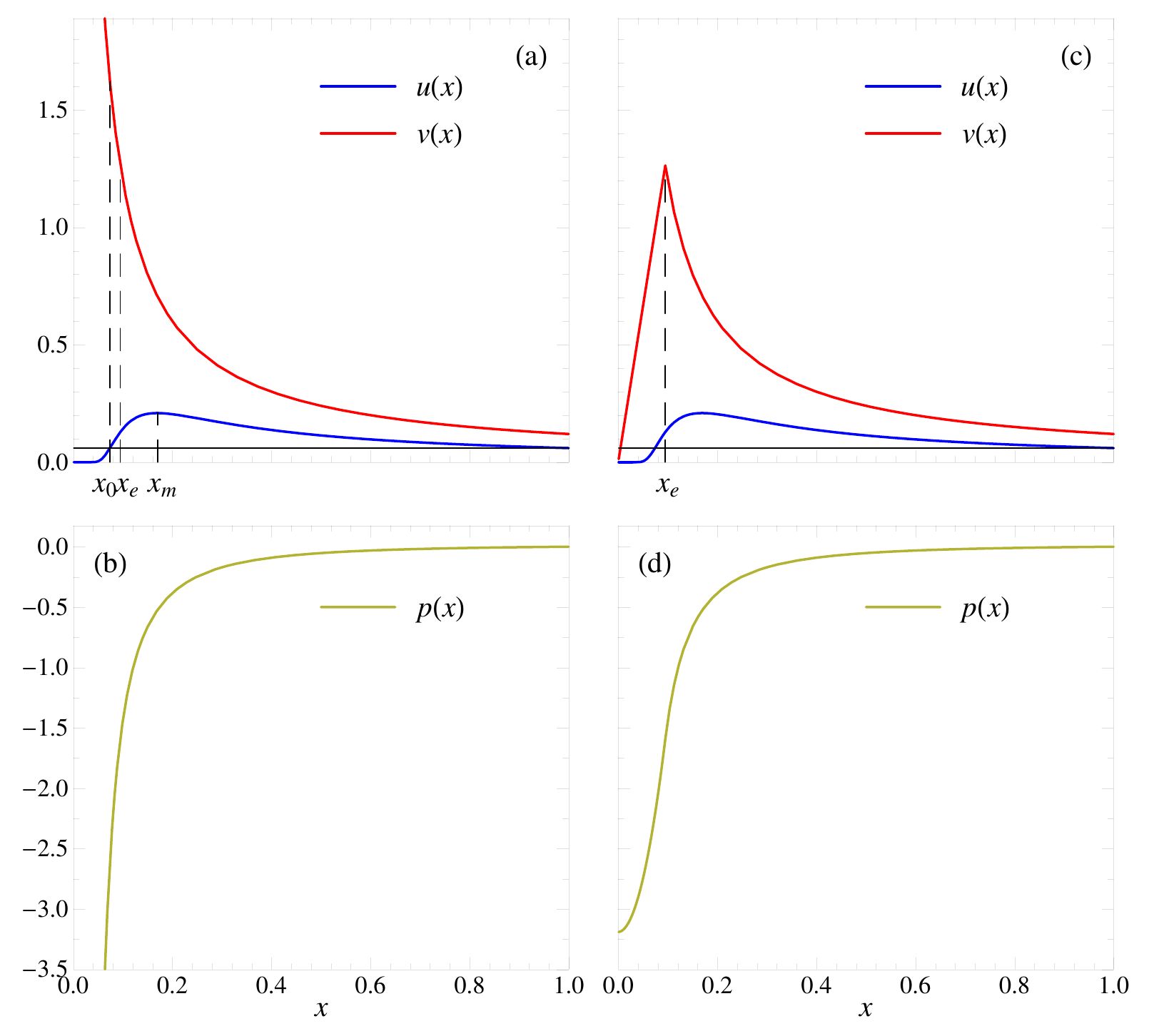}
\end{center}
\caption{Dimensionless radial $u$ (\ref{us}) and tangential $v$ (\ref{v2}) velocities
and pressure fall $p(x)$ (\ref{p2}) as dependent on radius to the hurricane center $x$.\newline
(a), (b): velocities and pressure assuming the eye is motionless;\newline
(c), (d): velocities and pressure with an account of eye rotation;\newline
$x_0$ is eye radius (the eye is motionless);\newline
$x_e$ is windwall radius (the eye is rotating);\newline
$x_m$ is the point where radial velocity reaches its maximum;\newline
$a = r_p\omega/u_c$ is the dimensionless angular momentum that is conserved on the streamline
outside the eye; $r_p$ is the outer radius of the hurricane streamline; $u_c = (2\Delta p/\rho)^{1/2}$ is
the condensational velocity scale, see (\ref{var1}) and notes to Fig.~4c; $\Delta p = \gamma p$ is
pressure drop due to condensation, $\gamma \equiv p_v/p$ is the relative share of water vapor, $p$
is air pressure outside the hurricane.\newline
The curves correspond to the following parameters: $r_p = 400$~km, $\omega = \Omega \sin\vartheta =
2.5\times 10^{-5}$~s$^{-1}$, $\vartheta = 20^{\rm o}$, $\gamma = 0.042$, $T = 30^{\rm o}$~C,
$\rho =1.22$~kg~m$^{-3}$, $p = 10^5$~J~m$^{-3}$, $a = 0.12$, $u_c = 83$~m~s$^{-1}$.
The black horizontal line in panels (a) and (c) denotes $u_1 = 0.06$.
}
\end{figure}

Note that the two branches of pressure $p(x)$ (\ref{p2}) inside and outside the eye join smoothly at $x = x_e$, while
the derivative $\pt p/\pt x$ features a minor discontinuity at this point. This discontinuity is practically unnoticeable in Fig.~5d.
This discontinuity has a clear physical meanining, but it is difficult to observe it at small $a$.
If using the smallness of $a$, $x_e$ and $x_m$ one chooses $x_e = x_m$ at the point where radial velocity has its maximum, then
the derivatives $\pt p/\pt x$ also coincide in this point, see (\ref{x0}). This is a consequence
of Eq.~(\ref{us}) and of the fact that $\pt u/\pt x = 0$ at $x = x_m$ \citep{mg09b}.

Equation~(\ref{x0}) represents an equation on the value of $x_0$ as a function of $a$ (\ref{xe}), $x_0=x_0(a)$, Fig. 4a.
The value $x_0 = x_m = 0.6$ is reached at $a = 0.43$, where $x_0^2 = 0.36$ and the approximation ({\ref{x0}) still holds.
Further widening of the eye radius $x_0$ involves the second branch of the solution of Eq.~(\ref{x0}) shown
in Fig.~4a. For this branch the growth of $x_0$ to $x_0 = 1$ corresponds to decreasing $a$.
On this branch potential energy arising from condensation that occurs
in the area $x_0 \le x \le 1$ and disappears at $x_0 = 1$, $a = 0$, should have been spent on rotation of a very wide
eye at low tangential velocities, i.e., the hurricane cannot form.

Transition from this branch with $x_0 > x_m$ to the branch $x_0 < x_m$ corresponds to formation of a hurricane.
In Fig.~4c the areas of possible maximum tangential velocities are shown for several values of $r_p\sin\vartheta$ as dependent on oceanic surface
temperature $T_s$ that dictates the value of condensational velocity $u_c$ (\ref{var1}).
We note two peculiarities: (1) temperature that corresponds to the transition between the two regimes
decreases with diminishing $r_p\sin \vartheta$ and (2) for any given temperature, the difference
between velocities of the two branches grows with decreasing $r_p\sin\vartheta$.
In the equatorial zone where $\sin\vartheta \to 0$, at observed surface temperatures
the difference between velocities on the two branches tends to infinity. Hurricanes cannot form.

As one can see from Fig.~4c, for any given temperature, the distance between the branches increases (while the probability of jumping from one branch to another consequently
decreases) with decreasing radius and/or latitude. For example, for curve 5 at 300~K the distance between the green and red branches
is nearly 80~m~s$^{-1}$, while for curve 2 it is about two times smaller.
The transition point between the two branches for curve 5 corresponds to $T = -5^{\rm o}$C, which is a temperature
when the hurricanes do not form. The transition point moves to the region of lower temperatures with growing $r_p\sin\vartheta$.
On the other hand, with increasing radius and/or latitude
the area of allowed velocities becomes narrower. For example, for curve 1 at the observed temperatures only a narrow range of
velocities in the vicinity of 46~m~s$^{-1}$ can be developed. These physical limitations shape the observed distributions of
hurricane wind velocities over temperature, Fig.~4c.

Finally, when the angular momentum $a$ decreases (due to either decreasing radius $r_p$ of
the condensation area as well as due to decreasing angular velocity $\omega$ at low latitudes), the eye radius
decreases as well. The energy ot the eye rotation, which in the dimensionless variables (\ref{var2})
is equal to $a^2/4$, see~(\ref{seg}), becomes small. In the result, there appears a  possibility for
tangential velocity in the windwall to grow to catastrophic values that are characteristic of tornadoes.

\section{Angular momentum: the effect of superposition of several streamlines}

Consideration of the energy budget of every streamline starts from $r = r_p$
and ends with $r = 0$, with the account made for the substraction of energy
to sustain the eye rotation. Eye radius $x_e$ for each streamline depends
on angular momentum $a = r_p\omega/u_c$, i.e., it is also determined
by the value of $r_p$. Angular momentum of the rotating air volumes in the eye
changes with distance according to $ar^2/r_e^2 = a x^2/x_e^2$. Thefore,
the complete information about a given streamline is contained in the
value of angular momentum $a$ that is conserved in the region outside
the eye $x \ge x_e$.

Hurricanes apparently comprise a large number of individual streamlines,
some of which can be easily traced on the satellite images as individual "tails"
spiraling towards the hurricane center -- the eye. These
streamlines can start at various radii $r_p$ and possess various values
of angular momentum $a$ that is conserved outside the eye.

The average angular momentum of the hurricane can be obtained considering the
sum of kinetic energies of rotation for all streamlines
as dependent on distance $r$ to the circulation center.
For each $i$-th
streamline starting at $r = r_{pi}$ the dependence of angular momentum $a_i(x)$ and
tangential velocity $v_i(x)$ on relative distance $x$ has the form
\beq
\label{ai}
\begin{split}
a_i(x) = a_i \left[\frac{x^2}{x_{ei}^2}\vartheta(x_{ei}-x) + \vartheta(x-x_{ei})\vartheta(x_i-x))\right],\,\,\,v_i(x) = \frac{a_i(x)}{x},\\
x_i\equiv \frac{r_{pi}}{r_{p\,max}},\,\,x \equiv \frac{r}{r_{p\,max}} \le 1,\,\,a_i=\frac{r_{pi}\omega}{u_c},\,\,
\vartheta(X) \equiv \begin{cases} 1, & \mbox{if } X \ge 0 \\
0, & \mbox{if } X <0 \end{cases}.
\end{split}
\eeq
\noindent
Here $x$ is defined relative to the streamline with maximum ultimate radius $r_{p\,max} \ge r_{pi}$
and maximum conserved angular momentum; $\vartheta(X)$ is the step function. Mean angular momentum $a(x)$ as dependent on $x$
can be obtained from the following relationship:
\beq
\label{ax}
a(x) = \frac{\sum_i{a_i(x)v_i^2(x)}}{\sum_i{v_i^2(x)}}=\frac{\sum_i{a_i^3(x)}}{\sum_i{a_i^2(x)}} \approx \frac{1}{n}\sum_i{a_i(x)},
\eeq
\noindent
where $a_i(x)$ and $v_i(x)$ are defined in (\ref{ai}). The last approximate equality in (\ref{ax}) is
written from the condition $\sum_i{a_i^2(x)} \approx na^2(x)$, $a_i^3(x) \approx a^2(x) a_i(x)$
to qualitatively evaluate the behavior of $a(x)$. As one can see from Eq.~(\ref{ax}),
the $a_i$ values that enter the sum diminish with decreasing $x$. Hurricane eye acquires a complex structure
that is formed by different streamlines at different radii $x_{ei}$. The closer to the center,
the smaller values of angular momentum are that form the hurricane eye. These two facts testifiy that,
in agreement with observations, $a(x)$ decreases towards the center despite the angular momentum
is conserved within each particular streamline.

\section{The pole approximation}

The circulation system of hurricanes and tornadoes can be considered in the pole approximation.
Such a consideration preserves the main physical features and the quantitative estimates
of velocities and condensational potential. In the pole approximation one preserves the terms
that display most singularity at $x \to 0$, namely, the major powers $1/x$ and $\ln x$. This is
equivalent to discarding $\pt u/\pt x$ as being small compared to $u/x$ in the Bernoulli integral (\ref{u})
and in the Euler equation for the streamline that corresponds to differentiating both parts of (\ref{u}) over $x$. In this case
the Bernoulli integral (\ref{u}) simplifies to become \citep{mg09b}
\beq
\label{BIp1}
u^2 + v^2 + w^2 + \ln x =0,\,\,\,v^2 = \frac{a^2}{x^2},\,\,\,w^2 = \beta^2\frac{u^2}{x^2},
\eeq
\beq
\label{BIp2}
u^2 = \left(-\ln x -\frac{a^2}{x^2}\right)\frac{x^2}{x^2 + \beta^2},\,\,\,\beta \equiv \frac{h_\gamma}{r_p},
\,\,\,p(x) = \ln x.
\eeq

For hurricanes, where $\beta \ll 1$, Eq.~(\ref{BIp2}) takes the form
\beq
\label{hur}
u^2 = -\ln x -\frac{a^2}{x^2}.
\eeq
\noindent
The point of maximum $x_m = \sqrt{2}a$. For maximum radial velocity $u_m^2$ we have
\beq
\label{um}
u_m^2 = -\ln(\sqrt{2}a)-\frac{1}{2}.
\eeq
\noindent
Velocities $u$ and $w$ become strictly equal to zero at $x = x_0$ defined from the equation
\beq
\label{x0p}
-\ln x_0 - \frac{a^2}{x_0^2} = 0.
\eeq
\noindent
Tangential velocity reaches its maximum $v_e = a^2/x_e^2$ at $x_e = 1.28 x_0$.

As one can see from the comparison of (\ref{BIa}) and (\ref{BIp1}), the pole approximation
consists in replacing condensational potential $\ln [(u/u_1)x]$ for $\ln x$, i.e., it does
not depend on the initial velocity $u_1$. The gradient of the potential $1/x +
u^{-1}\pt u/\pt x$ is replaced by $1/x$, i.e. the term containing $\pt u/\pt x$ is discarded.
This changes
the profile of the potential in the region from $x = 1$, $u = u_1 = u(1)$ to
$x = x_0$, $u = u_0 = u(x_0)$, while the change of the potential (pressure difference $\delta p$)
is not affected by the approximation. As far as with decreasing $x$ the value of $u$
first grows at $x \ge x_m = \sqrt{2}a$ and then declines at $x \le x_m$, the exact potential declines
more slowly than the potential in the pole approximation at $x \ge x_m$ and more
rapidly than the latter at $x \le x_m$.

Tornadoes, where $\beta \ge 1$, preserve the main features of the pole approximation. Angular momentum in the tornado
is very small, $a = r_p\omega /u_c \ge h_\gamma\omega/u_c \le 0.004$ (at $h_\gamma \sim 10$~km,
$\omega \sim 3\times 10^{-5}$~s, $u_c \sim 80$~m~s$^{-1}$).
Due to the fact that the eye in tornadoes is extremely narrow,
$r_0 = h_\gamma x_0 = h_\gamma a/2 \sim 20$~m, the eye rotation practically does not take any energy
from the streamline. Neglecting the difference between $x_0$ and $x_e$
we have from Fig.~4a that $x_0 \sim x_0 \sim a/2$, see Fig.~6a. From (\ref{BIp2})
we have $\delta p \sim p(1) - p(x_0) = -\ln x_0 \sim 6$, Fig.~6b, so that
$\delta p\Delta p/p \sim 6 \gamma \sim 0.3$. In the center of the tornado
the air pressure can drop by 30\% compared to the outer environment.
This causes tangential velocity
to grow to very high values  of the order of $(a/x_0)u_c \sim 2u_c \sim 160$~m~s$^{-1}$, Fig.~6a.
In contrast to the hurricane, while approaching $x = x_0$ the radial velocity $u$
in tornado decreases proportionally to $x^2$, see~(\ref{BIp2}) and Fig.~6a, and has a maximum
at $x = 0.69$ ($-\ln x = 1/2$.). Vertical velocity
$w$ (\ref{BIp1}) in tornado has the same behaviour as radial velocity $u$ in the hurricane,
see (\ref{BIp2}). In the point of maximum $x_m = \sqrt{2}a$ the vertical velocity reaches $w_m =
u_c(-ln(\sqrt{2}a)-1/2)^{1/2} \sim 160$~m~s$^{-1}$, see~(\ref{um}). This maximum
value of vertical velocity coincides with the value of tangential velocity
reached at $x_0 \sim a/2$, Fig.~6a. Radial velocity $u$ in tornadoes
changes much more slowly at small $x$ than it does in hurricanes, $\pt u/\pt x \ll
u/x$, which makes the pole approximation more exact for tornadoes than it is for the hurricanes.

\begin{figure}
\begin{center}
\includegraphics[width=0.9\textwidth]{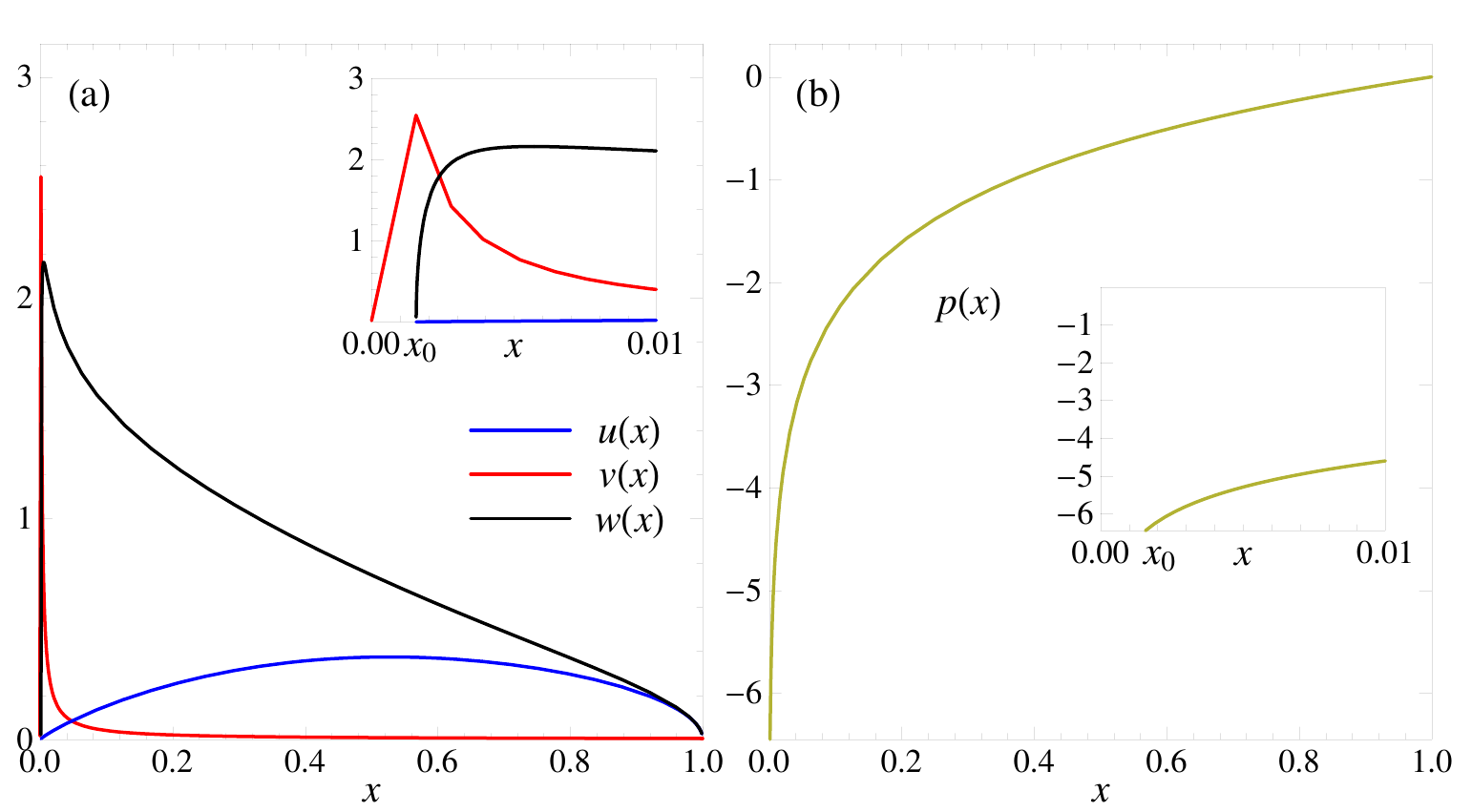}
\end{center}
\caption{Wind velocities and pressure profile in a tornado calculated in the pole approximation.\newline
(a): vertical $w$, tangential $v$ and radial $u$ velocities in units of $u_c = (2\Delta p/\rho)^{1/2}$
(\ref{var1}) depending on distance $x$ to the condensation center ($x$ measured in units of
the condensational scale height $h_\gamma$ (\ref{grp}) at $\beta = 1$) with dimensionless angular momentum
$a = h_\gamma\omega/u_c$, $x_0$ is the eye radius that satisfies $\ln x_0 = a^2/x_0^2$,
Typical values at $30^{\rm o}$C are $u_c \sim 80$~m~s$^{-1}$,
$h_\gamma \sim 10$~km, Fig.~2, $\Delta p \sim 4 \times 10^3$~Pa, $a = 0.004$, $x_0 = 0.0016$.\newline
(b): pressure fall $p(x)$ in units of $\Delta p$ from the outer edge of the condensation area at $x = 1$ to
the windwall -- eye radius at $x = x_0$.}
\end{figure}

We note one more peculiarity of the tornado. The eye radius is $x_0 \sim a/2$, Fig.~4,
$a =h_\gamma \omega/u_c$, $u_c = (2\Delta p/\rho)^{1/2} = (2p/\rho)^{1/2}\gamma^{1/2}$, see (\ref{p}).
The water vapor relative partial pressure $\gamma$ decreases approximately
exponentially with growing height and practically turns to zero at 5--8~km, see~(\ref{1LT}).
Therefore, the eye radius in the tornado grows proportionally to $\gamma^{-1/2}$ with increasing
height $z$. This creates a conspicuous "mouth" of the tornado funnel -- the funnel is narrow
at the surface and widens as the height grows. This is another indication that tornado represents
a single streamline in contrast to the hurricane, where the eye is formed as the result
of many streamlines being superimposed.

\section{Major features of atmospheric circulation induced by water vapor condensation}

1. The physical cause of the condensation-induced circulation consists in the fact that hydrostatic
equilibrium of the moist terrestrial atmosphere is unstable and cannot exist (Sections 1, 2).

In an isothermal atmosphere partial pressure of water vapor would exponentially decline with
height decreasing twofold per each nine kilometers of height increment.
According to Clausius-Clapeyron law, saturated water vapor would follow the same exponential distribution
if the air temperature dropped by approximately ten degrees~K  per each nine kilometers of height increment,
i.e., with a lapse rate of 1.2~K~km$^{-1}$. However, adiabatic ascent of a volume of moist air causes air temperature
to drop at a significantly higher rate bounded between approximately 4 and 9.8~K~km$^{-1}$.
Consequently, a random adiabatic displacement of moist
air upwards makes the excessive water vapor condense and leave the gas phase. Pressure in the adiabatically
ascending air drops. There appears an upward-direction evaporative-condensational force that sustains
continuous adiabatic ascent of the moist air and causes air to circulate in the vertical and horizontal directions.

2. Stationarity of the annually averaged state of the atmosphere and the surface testify to the
equality between the global mean rates of evaporation and condensation rates. However, evaporation
is maintained by the stationary flux of solar radiation. Mean power of the evaporation flux cannot
exceed the power of solar radiation. In contrast, the power of the flux of water vapor condensation is not related to the
solar radiation power and can locally exceed the evaporation power by one-two orders of magnitude
(e.g., in hurricanes and tornadoes) or become by the same factor less intense than local
evaporation (in the regions of the anticyclonic descent of air masses). These contrasting
properties of evaporation and condensation are crucial for the condensation-induced circulation.

3. In a large-scale horizontal circulation the air flows from the donor area to the
acceptor area in the lower atmosphere and in the opposite direction in the upper atmosphere.
In the donor region the air masses descend; there is no condensation, but the evaporation is present.
In the acceptor area the air masses ascend; there is condensation; both
locally evaporated water vapor and the water vapor imported from the donor area undergo condensation.
The rate of condensation is thus always higher in the acceptor area than in the donor area;
this sustains the pressure gradient necessary for the maintenance of the air flow.
Water returns to the donor area in the liquid phase. In a large-scale circulation total evaporation
can coincide with total condensation (then the circulaton can be stationary in the long-term).
However, the circulation itself is only possible due to the absence of equality between
local evaporation and condensation (they do not coincide in either donor or acceptor area). Condensation
rate in the acceptor area must be higher than in the donor area. This is stably achieved by higher
evaporation rate in the acceptor area compared to the donor area.

4. In large-scale (Hadley, biotic pump of continental-scale natural forest)
and medium-scale (moonsoon) circulation systems friction forces prevent significant
acceleration of air masses and do not allow high wind speeds to develop.
If there is no control over condensation and evaporation, the areas where such
circulation systems develop are also prone to the appearance of frequent floods,
droughts, tornadoes and hurricanes of varying intensity. The condensation-induced
circulation can be regulated only with use of the genetically based biological
programs of natural forest ecosystems, where the necessary soil moisture store
is continuously maintained by biotic regulation of evaporation and condensation.

5. Compact circulation takes place where friction is small. Condensation power
that locally exceeds the power of evaporation by many times accelerate the air
to catastrophic velocities observed in hurricanes and tornadoes. In a compact
circulation there is a pronounced condensation center, to which the rotating
air masses converge. The condensation center moves continuously in
the horizontal plane. This movement guarantees that the circulation system
is supplied by new amounts of water vapor which is depleted as the system functions.
If compact circulation do not move, they dissipate (similar to the movement
of an animal over its feeding territory).

6. With increasing mixing ratio $\gamma$ of water vapor in the moist air
and as this ratio approaches unity due to the removal of the non-condensable
air constituents (oxygen and nitrogen), the moist adiabatic lapse rate
approaches the hydrostatic value of 1.2~K~k,$^{-1}$. The evaporative-condensational
force that supports the adiabatic ascent tends to zero.
The adiabatic ascent stops; the atmosphere becomes vertically isothermal;
condensation-induced circulation systems of all types disappear. But with increasing
temperature the evaporative-condensational force grows exponentially.

7. The condensation-induced circulation arises on horizontally isothermal surface
and does not demand an external differential heating to arise.

\end{document}